\documentclass[aps]{revtex4}
\usepackage{amsfonts}
\usepackage{amsmath}
\usepackage{amssymb,epsf}
\usepackage{color}
\begin{document}

\title{Thermal stability of a special class of black hole solutions in $F(R) $ gravity}
\author{S. H. Hendi$^{1,2}$\footnote{
email address: hendi@shirazu.ac.ir}, R. Ramezani-Arani$^{3}$, E.
Rahimi$^{3}$} \affiliation{$^1$ Physics Department and Biruni
Observatory, College of Sciences, Shiraz
University, Shiraz 71454, Iran\\
$^2$ Research Institute for Astronomy and Astrophysics of Maragha
(RIAAM), P.O. Box 55134-441, Maragha, Iran\\
$^{3}$ Department of Elementary Particles, Faculty of Physics,
University of Kashan, Kashan, Iran}

\begin{abstract}
In this paper, we work on the topological Lifshitz-like black hole
solutions of a special class of vacuum $F(R)-$gravity that are
static and spherically symmetric. We investigate geometric and
thermodynamic properties of the solutions with due respect to the
validity of the first law of thermodynamics. We examine the van
der Waals like behavior for asymptotically AdS solutions with
spherical horizon by studying the $P-v$, $G-T$ and $C_{Q,P}-r_{+}$
diagrams and find a consistent result. We also investigate the
same behavior for hyperbolic horizon and interestingly find that
the system under study can experience a phase transition with
negative temperature.
\end{abstract}

\maketitle

\section{Introduction}

$F(R)$ gravity is one of the best models of modified general relativity with
a renewed interest in recent years. In addition to the simple and general
Lagrangian of this model, the main (but not the first) motivation of
considering $F(R)$ gravity with arbitrary function of Ricci scalar is that
one can explain the accelerated expansion and structure formation of the
Universe without considering dark energy or dark matter. It is also believed
that some curvature corrections arising from quantum theory of gravity may
be collected to special functional forms of $F(R)$ gravity.

In other words, some defects in Einstein's general relativity and the
motivation of studying more complete and general gravitational models, have
led to the creation of general modified gravity models. These gravitational
models have been considered in various branches of gravity, cosmology,
astrophysics and their interesting results have made these models
appropriate generalization for the Einstein's gravity \cite%
{Cognola,Cooney,Guo}. In the lastest decade, special attention was given to
modified theories that generalize the gravitational action integral with the
simplest model, the so-called $F(R)$ gravity \cite{Sotiriou2}. There are two
approaches for obtaining the field equations of the generalized $F(R)$
gravity. The first one is the standard metric formalism and the other one is
the Palatini method \cite{Palatini} in which metric tensor and affine
connection are treated, a priori, as independent variables. There are many
viable $F(R)$ models that can satisfy both cosmological and local gravity
constraints \cite{Amendola,Li,Hu,Starobinsky}. Moreover, $F(R)$ gravity can
be reduced to general relativity in the specific cases \cite{Capozziello1}.
Nowadays, $F(R)$ gravity has attracted much attentions of the researchers
and there are many studies in this field \cite%
{Hendi1,Hendi2,Capozziello2,Maeda}.

The generalization of $F(R)$ theory in the context of Horava--Lifshitz
gravity and its interesting cosmological results have been addressed in
series of papers \cite%
{F(R)-HL1,F(R)-HL2,F(R)-HL3,F(R)-HL4,F(R)-HL5,F(R)-HL6,F(R)-HL7,F(R)-HL8,F(R)-HL9,F(R)-HL10}%
. The pioneer work of Horava in $2009$ \cite{Petr}, the so-called
Horava--Lifshitz gravity is one of the candidate theories of quantum
gravity. Since general relativity is not a renormalizable theory (it means
that it is successful as a classical theory of gravity, but it breaks down
at some scale), and therefore, it should be viewed as an effective theory.
Beyond that scale, general relativity is not suitable theory to describe the
gravitational interactions or spacetime itself and one cannot construct its
quantum counterpart using conventional quantization techniques.

On the other hand, within this perspective that general relativity is an
effective theory, it may solve some problems in gravitation by treating the
quantum concept more fundamentally, and so, space and time are not
equivalent (anisotropic) at high energy level. The relativistic concept of
time with its Lorentz invariance emerges at large distances. The theory
relies on the theory of foliations to produce its causal structure. It is
related to topologically massive gravity and the Cotton tensor. So it may
have a possible UV completion of general relativity to address this issue.
Recently, the proposed Horava-Lifshitz gravity promises a UV completion of
Einstein's theory by sacrificing general covariance at short distances and
introducing anisotropic spacetime scaling. In addition, Donoghue showed that
general relativity and quantum field theory can be perfectly compatible if
the quantum gravity is formulated as an effective field theory \cite%
{Donoghue}. Therefore, the effective field theory of quantum gravity is
valid above which the effective description is replaced with the UV
completion. Furthermore, Horava-Lifshitz gravity theory can be employed as a
covariant framework to build an effective field theory for the fractional
quantum Hall effect that respects all the spacetime symmetries such as
non-relativistic diffeomorphism invariance and anisotropic Weyl invariance
as well as the gauge symmetry. Consequently, investigation of
Horava-Lifshitz gravity family can help us to deepen our insight for moving
from classical gravity point of view to quantum one.

According to the discovery of Hawking radiation \cite{Hawking1}, one finds
that the black hole thermodynamics has crucial role for studying the quantum
nature of gravity. Hawking tried to show that black holes behave like black
bodies in the usual thermodynamic sense, emitting radiation with a thermal
spectrum \cite{Hawking1,Hawking2}. In this regard, many authors have tried
to obtain a new approach to find a statistical origin to Bekenstein-Hawking
entropy \cite{Strominger-Vafa}.

The entropy of Einsteinian black hole, known as the Bekenstein-Hawking area
law entropy, suggests that the quantum degrees of freedom of a typical black
hole are effectively distributed over a surface, rather than a volume.
Following the works of 't Hooft \cite{Hooft} and Susskind \cite{Susskind},
one can find this crucial result based on the holographic principle, in
which says that quantum gravity in a given volume should be described by a
theory on the boundary of that volume. The mentioned holographic principle
is related to the so-called anti-de Sitter/conformal field theory (AdS/CFT)
correspondence \cite{Maldacena,Aharony}. In the other words, in order to
explain the holographic principle, a consistent quantum gravity theory
should admit two equivalent descriptions (AdS/CFT correspondence): one as a
bulk semiclassical general relativity theory, and one as a boundary quantum
field theory. So, investigation of black hole solutions with AdS asymptote,
which has an undeniable role for constructing a consistent quantum gravity
theory \cite{AdS-QG}, has special interest.

Recently, a renewed interest in phase transition of asymptotically AdS black
holes, especially van der Waals like \cite{Dolan}, has appeared.
Subsequently, a number of interesting results, such as, triple points \cite%
{Altamirano}, reentrant phase transitions \cite{Gunasekaran}, and analogous
Carnot-cycle heat engines \cite{Johnson} were obtained. Regardless of
massive gravity black holes \cite{PRDrapid}, it is known that the van der
Waals behavior in Einstein gravity is seen only for AdS black holes with
spherical horizon \cite{Dolan}, and therefore, such behavior does not take
place for AdS black holes with flat or hyperbolic horizons (no real critical
point is found). In this paper, we observe that in addition to spherical
horizon, one can find the van der Waals behavior in the especial case of $%
F(R)$ gravity with hyperbolic horizon. It is note that in this case,
although the critical volume and pressure are positive, the critical
temperature is negative. Since negative temperature has a physical
interpretation in usual thermodynamics of quantum system (e.g: a system of
nuclear spins in an external magnetic field or population inversion in laser
\cite{negativeT}, it will be interesting to investigate our solutions with
quantum mechanical point of view.

Thermodynamic descriptions of gravitational solutions at the event horizon
and cosmological ones at the apparent horizon in $F(R)$ gravity have been
studied in series of papers \cite%
{F(R)-Thermo1,F(R)-Thermo2,F(R)-Thermo3,F(R)-Thermo4,F(R)-Thermo5,F(R)-Thermo6,F(R)-Thermo7,F(R)-Thermo8,F(R)-Thermo9}%
. In this work, we are going to evaluate thermodynamic phase transition of
black hole solutions in a special class of $F(R)$ gravity with constant
Ricci scalar in which both $F(R)$ and its derivative $F_{R}(R)$ are vanished
in the field equations. The structure of our paper is as follows: in next
section, we are going to introduce $F(R)$ gravity, in brief, and then we
obtain the special class of exact solutions with black hole interpretation.
Then we investigate geometric and thermodynamic properties. In Sec. \ref%
{Extended}, we work in the extended phase space and investigate the possible
phase transition and van der Waals like behavior. We also discuss thermal
stability and critical quantities. Final section is devoted to some
concluding remarks.

\section{Basic Equations and Dynamic Black hole solutions}

\label{FieldEq}

The purpose of this paper is to study the thermodynamics of a typical
anti-de Sitter black hole solution in four dimensional $F(R)=R+f(R)$ gravity
in which the Ricci scalar is constant $(R=R_{0}).$ Let us first consider the
$4$-dimensional action of $R+f(R)$ gravity which is given by
\begin{equation}
S=\int_{\mathcal{M}}d^{4}x\sqrt{-g}F(R)=\int_{\mathcal{M}}d^{4}x\sqrt{-g}%
[R+f(R)],  \label{Action}
\end{equation}%
where $\mathcal{M}$ is a four-dimensional bulk manifold. It is clear that
for $F(R)=R$ ($f(R)=0)$, one can recover the Hilbert-Einstein action of
General Relativity. Using the variational principle on the action of $F(R)$
gravity (\ref{Action}), it is a matter of calculation to show that the field
equation is given by
\begin{equation}
R_{\mu \nu }[1+f_{R}]-\frac{1}{2}[R+f(R)]g_{\mu \nu }+[g_{\mu \nu }\square
-\nabla _{\mu }\nabla _{\nu }]f_{R}=0,  \label{Field
equation}
\end{equation}
where we use the notation $A_B=\frac{dA}{dB}$. It is easy to show that one
can rewrite Eq. (\ref{Field equation}) with the following form
\begin{equation}
G_{\mu \nu}F_{R}=\frac{1}{2}g_{\mu \nu}[F(R)-RF_{R}]+[\nabla _{\mu }\nabla
_{\nu }-g_{\mu \nu}\square]F_{R}.  \label{FE2}
\end{equation}

In this paper, we follow the method of Ref. \cite{Marco}, which is a special
class of $F(R)$-gravity models with two constraints, simultaneously, $%
F(R_{0})=0$ and $F_{R}=0$. Taking into account the mentioned constraints,
one can find that the vacuum equation (\ref{FE2}) are automatically
satisfied with arbitrary $R_{0}$. It is notable that the mentioned class
does not cover the usual general relativity solutions since the vacuum field
equation of general relativity with arbitrary metric identically satisfied $%
F_{R}=1$ with vanishing Ricci scalar. Consequently, our strategy is working
on some viable models of $F(R)$-gravity that satisfy these conditions and
solving the equation of constant Ricci scalar. As it is mentioned in \cite%
{Marco}, there are several models for the early-time inflation or late-time
accelerated expansion that can satisfy the mentioned constraints. As a
result, we regard a spherically symmetric and static solutions with constant
Ricci scalar which are reported in \cite{Marco} to investigate their
possible phase transition.

Here, our main motivation is the study of thermodynamical and geometrical
aspects of topological black hole solutions with Lifshitz-like spacetime.
Therefore, we consider the metric of $4$-dimensional spacetime as \cite%
{Marco}
\begin{equation}
ds^{2}=-e^{2\alpha (r)}B(r)dt^{2}+\frac{dr^{2}}{B(r)}+r^{2}d\Omega ^{2},
\label{Metric}
\end{equation}%
in which
\begin{equation}
d\Omega_{k} ^{2}=\left\{
\begin{array}{cc}
d\theta ^{2}+\sin ^{2}\theta d\varphi ^{2} & k=1 \\
d\theta ^{2}+d\varphi ^{2} & k=0 \\
d\theta ^{2}+\sinh ^{2}\theta d\varphi ^{2} & k=-1%
\end{array}%
\right. ,
\end{equation}%
where $k=1$, $0$ and $-1$ represent spherical, flat and hyperbolic horizon
of possible black holes, respectively. Hereafter, we indicate $\omega_{k}$
as the volume of boundary $t=cte$ and $r=cte$ of the metric. Since we desire
to study the Lifshitz-like solutions \cite{Marco}, we define $\alpha(r)$ as
\begin{equation}
\alpha (r)=\frac{1}{2}\ln \left( \frac{r}{r_{0}}\right) ^{z},  \label{alfa}
\end{equation}%
where $z$ is a real number and $r_{0}$ is an arbitrary (positive) length
scale. It is notable that in order to obtain a dimensionless argument of
logarithmic function, the existence of $r_{0}$ is necessary. Inserting Eq. (%
\ref{alfa}) in the introduced metric (\ref{Metric}), one can obtain
\begin{equation}
ds^{2}=-(\frac{r}{r_{0}})^{z}B(r)dt^{2}+\frac{dr^{2}}{B(r)}+r^{2}d\Omega
^{2}.  \label{Metric2}
\end{equation}

As an additional comment, we note that although we can start with Eq. (\ref%
{Metric2}) as the line element, it is convenient to define Eq. (\ref{Metric}%
) at the first step to guarantee that the Lifshitz factor ($e^{2\alpha (r)}$%
) is positive definite and change of signature of the metric comes from the
sign of $B(r)$ (the number of plus ($+$) and minus ($-$) signs is
unchanged). Considering the metric (\ref{Metric2}) with the mentioned field
equation (\ref{Field equation}) (and also the mentioned constraints for
special class of $F(R)$-gravity), one can extract the metric function for $%
R=R_{0}$, where
\begin{equation}
R=R_{0}=-\frac{d^{2}B(r)}{dr^{2}}-\frac{3z+8}{2r}\frac{dB(r)}{dr}-\frac{%
z^{2}+2z+4}{2r^{2}}B(r)+\frac{2k}{r^{2}},  \label{Ricci}
\end{equation}%
with the following exact solutions \cite{Marco}
\begin{equation}
B(r)=K-\frac{C_{\pm }}{r^{b_{\pm }}}-\lambda r^{2},  \label{B(r)}
\end{equation}%
where $K$ and $\lambda $, are two (positive/negative or zero)
constants which their values are depending on the signs/values of
$k$ and $R_{0}$ as
\begin{equation}
K=\frac{4k}{z^{2}+2z+4},  \label{K}
\end{equation}%
\begin{equation}
\lambda =\frac{2R_{0}}{z^{2}+8z+24}.  \label{lambda}
\end{equation}

In addition, $C_{\pm }$ are two integration constants while $b_{\pm }$ is
\begin{equation}
b_{\pm }=\frac{1}{4}\left( 3z+6\pm \sqrt{z^{2}+20z+4}\right) .  \label{b}
\end{equation}

Avoiding complex values of $B(r)$, one may regard the following constrain on
$z$
\begin{equation}
z^{2}+20z+4>0\;\text{ or equivalently }\;z\notin \left[ -2(2\sqrt{6}+5),2(2%
\sqrt{6}-5)\right] .  \label{valid}
\end{equation}

According to the above constrain on $z$, one can find the allowed ranges of $%
b_{\pm }$ as follows
\begin{eqnarray}
b_{-} &\in &(-\infty , -13.35]\text{ \ \ \ \ \ \ \ \ \ when \ \ \ \ \ \ \ \
\ }z<-2(2\sqrt{6}+5), \\
b_{-} &\in &[0.93,+\infty )\text{\ \ \ \ \ \ \ \ \ \ \ \ \ \ \ when \ \ \ \
\ \ \ }z>2(2\sqrt{6}-5).
\end{eqnarray}
and
\begin{eqnarray}
b_{+} &\in &(-\infty , -12.93]\text{ \ \ \ \ \ \ \ when \ \ \ \ \ \ \ }%
z<-2(5+2\sqrt{6}), \\
b_{+} &\in &[1.35,+\infty )\text{ \ \ \ \ \ \ \ \ \ \ \ \ when \ \ \ \ \ \ \
} z> 2(2\sqrt{6}-5).
\end{eqnarray}

\begin{figure}[tbp]
$%
\begin{array}{cc}
\epsfxsize=7cm \epsffile{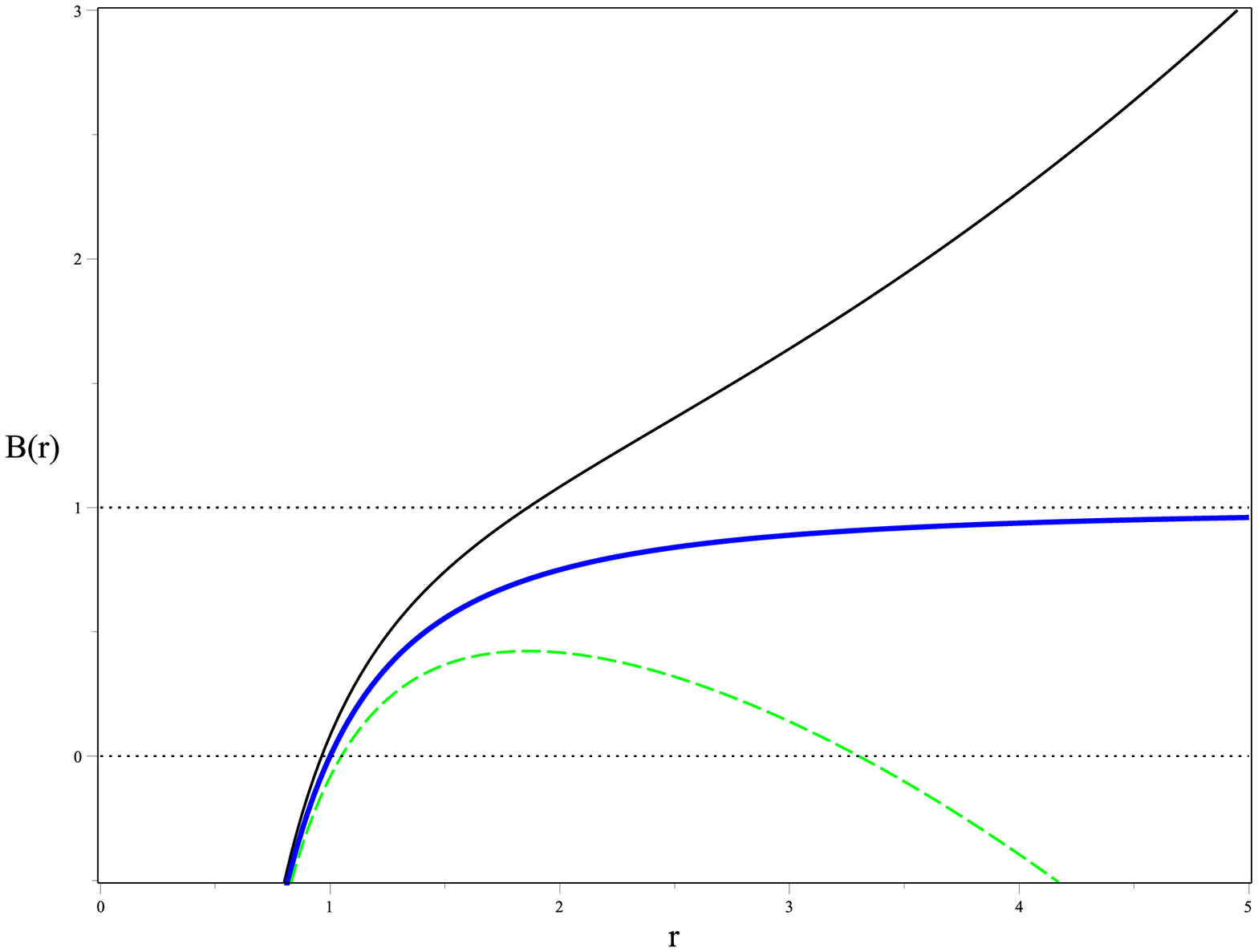} & \epsfxsize=7.5cm %
\epsffile{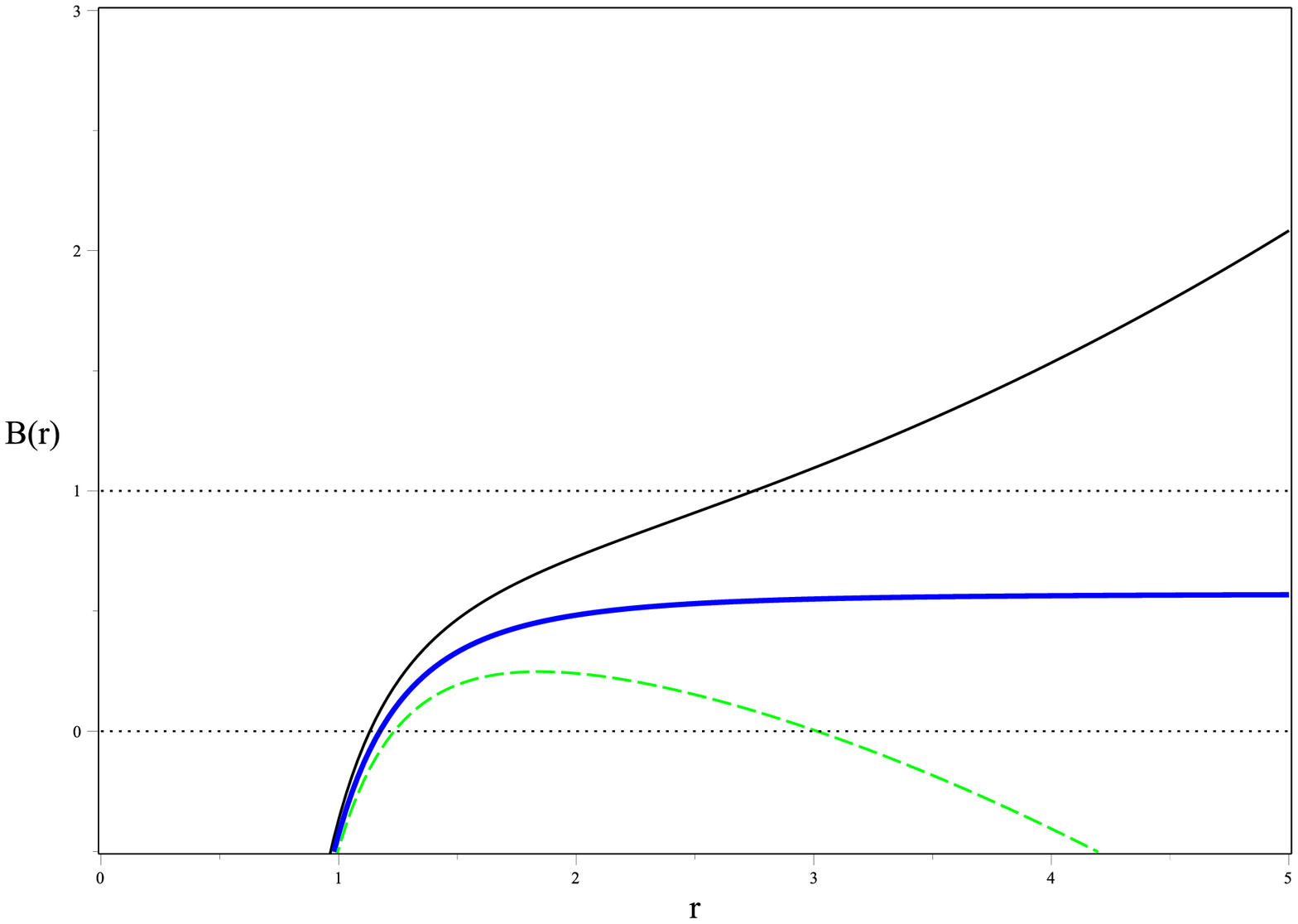}%
\end{array}
$%
\caption{ B(r) versus $r$ for $C_{+}=1$, $C_{-}=0$, $k=1$, and $R_{0}=-1$
(black-continuous line), $R_{0}=0$ (blue-bold line) and $R_{0}=1$
(green-dashed line). "\textbf{left figure: $z=0$ and right figure: $z=1$}" }
\label{FigB(r)}
\end{figure}

In order to interpret the solutions as black holes, we should examine the
existence of horizon and singularity for the singular black holes. The
presence of singularity could be investigated by studying curvature scalars
for which we choose the Kretschmann scalar. It is a matter of calculation to
show that for these solutions, the Kretschmann scalar is
\begin{eqnarray}
R_{\alpha \beta \gamma \delta }R^{\alpha \beta \gamma \delta } &=& \left(%
\frac{d^2B(r)}{dr^2}\right)^2+ \left(\frac{z\left( 3\frac{dB(r)}{dr}
+zB(r)-2B(r) \right) }{r^2}\right)\frac{d^2B(r)}{dr^2}+\frac{9z^2+16}{4r^2}
\left(\frac{dB(r)}{dr}\right)^2+  \notag \\
&& \left(\frac{z(3z^2-6z+8)B(r)}{2r^3} \right)\frac{dB(r)}{dr}+ \frac{%
z^4-4z^3+12z^2+16}{4r^4}B^2(r)-\frac{8k}{r^4}B(r)+\frac{4k^2}{r^4}.
\label{RR}
\end{eqnarray}

It is straightforward to show that Eq. (\ref{RR}) diverges at $r=0$ and it
is finite for $r \neq 0$. In addition, according to the Fig. \ref{FigB(r)},
one finds that the metric function has at least one real positive root (with
positive slope). As a result, the mentioned solutions can be interpreted as
black holes. In addition, according to Fig. \ref{FigB(r)} and also Eq. (\ref%
{B(r)}), one finds that for positive and negative $R_{0}$, one can find
asymptotically dS and adS, respectively with an effective cosmological
constant, $\Lambda_{eff}=\frac{6R_{0}}{z^2+8z+24}$. Moreover, it is clear
that for vanishing $R_{0}$ and $k=1$ the mentioned solutions are
asymptotically flat only for $z=0$ (we should note that although $z=-2$ with
$k=1$ leads to $K=1$, but $z=-2$ is not allowed based on Eq. (\ref{valid}) ).

Now, we are in a position to study thermodynamical properties of the
solutions and investigate their thermal stability based on the heat capacity.

\subsection{Conserved and thermodynamical quantities}

Here, we are going to calculate the conserved and thermodynamic quantities
of black holes in $F(R)$ gravity. Due to the fact that the employed metric
contains a temporal Killing vector, we use the concept of surface gravity to
calculate the temperature of black holes at the event horizon $r_{+}$
\begin{equation}
T=\frac{1}{2\pi }\sqrt{\frac{1}{2}(\nabla _{\mu }\chi _{\nu })(\nabla ^{\mu
}\chi ^{\nu })}=\frac{B^{\prime }(r)}{4\pi }\left( \frac{r}{r_{0}}\right) ^{%
\frac{z}{2}}|_{r=r_{+}},  \label{Temp}
\end{equation}%
where $\chi ^{\nu }=\delta _{0}^{\nu }$ is the Killing vector. Regarding Eq.
(\ref{Temp}) with metric function (\ref{B(r)}), one can find
\begin{equation}
T=\frac{\left[ (K-\lambda r_{+}^{2})b_{-}-2\lambda r_{+}^{2}\right]
r_{+}^{b_{+}}-C_{+}\left( b_{+}-b_{-}\right) }{4\pi r_{+}^{1+b_{+}}}\left(
\frac{r_{+}}{r_{0}}\right) ^{\frac{z}{2}},  \label{Temp2}
\end{equation}%
where $C_{+}$ is removed due to the fact that the metric function vanishes
on the event horizon, $r_{+}$.

In order to study the entropy of black holes in $F(R)=R+f(R)$ gravity, one
can use the generalized area law \cite{Cognola}. According to the result of
Ref. \cite{Cognola}, the entropy can be calculated as
\begin{equation}
S=\frac{A}{4}[1+f_{R}(R_{0})],  \label{Entropy}
\end{equation}%
where $A$ is the event horizon area of the black holes. But here we have
used two constraints $F(R)=0$ and $F_{R}=0$, and therefore, Eq. (\ref%
{Entropy}) leads to zero entropy. This problem comes from the fact that the
entropy of $F(R)$ gravity (in the non-equilibrium description of
thermodynamics \cite{ENT}) is a modification of the area law with an
effective gravitational coupling ($G_{eff}=G/F_R$). Since in our case we
have $F_R=0$ and the effective gravitational coupling diverges, we could not
use the mentioned modified area law relation. In other words, the usual Wald
approach is break down and we have to use an alternative method. Since we
believe that the black hole solutions should satisfy the laws of
thermodynamics, we use the first law to calculate the nonzero entropy.

Taking into account the timelike Killing vector $\left( \partial /\partial
t\right) $, one can show that the finite mass per unit volume $\omega _{k}$
can be obtained as
\begin{equation}
M=\frac{1}{2}\left( Kr_{+}^{b_{-}}-\lambda r_{+}^{2+b_{-}}-\frac{C_{+}}{%
r_{+}^{b_{+}-b_{-}}}\right) .  \label{Mass1}
\end{equation}

Here, we desire to calculate the entropy in by using the validity of the
first law of thermodynamics. It is easy to show that
\begin{equation}
\delta M=T\delta S,  \label{first}
\end{equation}%
and therefore, it is a matter of calculation to show that the following
equality holds
\begin{equation}
S=\int \frac{dM}{T}=\frac{4\pi r_{+}^{1+b_{-}}}{2-z+2b_{-}}\left( \frac{r_{0}%
}{r_{+}}\right) ^{\frac{z}{2}}.  \label{TU}
\end{equation}

It is notable that the obtained relation for the entropy reduces to the area
law for $z=0$ ($b_{-}=1$). In other words, it seems that the $F(R)$ gravity
does not direct effect on the entropy relation, such as that occurs in the
equilibrium description of thermodynamics in $F(R)$ gravity \cite{ENT}.

\section{Extended phase space thermodynamics, thermal stability and phase
transition}

\label{Extended}

Regarding the variation of the cosmological constant as the vacuum
expectation value of a quantum field, one may expect to consider it and its
conjugate in the first law of thermodynamics \cite{VaryLambda}. In this
regard, the cosmological constant interpreted as a dynamical pressure of the
black hole system as \cite{Dolan} (Note: $\lambda \approx \frac{\Lambda }{3}$%
)
\begin{equation}
P=-\frac{3\lambda }{8\pi },  \label{P}
\end{equation}%
where its conjugate extensive quantity is the thermodynamic volume which can
be obtained by%
\begin{equation}
V=\left( \frac{\partial H}{\partial P}\right) _{S,C_{+}},  \label{V}
\end{equation}%
in which $H$ is the enthalpy of system. We should note that in the extended
phase space the mass of black hole is not the internal energy of the system,
but its enthalpy $H\equiv M$. As an additional comment, it is worthwhile to
mention that the modified Smarr relation which can be calculated by the
scaling argument for our Lifshitz like solutions in the extended phase space
is%
\begin{equation}
M=\frac{(2b_{-}+2-z)}{2b_{-}}TS-\frac{2}{b_{-}}PV+\frac{b_{+}}{b_{-}}%
\mathcal{C}C_{+},  \label{Smarr2}
\end{equation}%
where $\mathcal{C}=\left( \frac{\partial M}{\partial C_{+}}\right) _{S,P}=%
\frac{-1}{2r_{+}^{b_{+}-b_{-}}}$ can be interpreted as a modified potential
per unit charge that calculated at the event horizon of Lifshitz like black
hole solutions. One can confirm that Eq. (\ref{Smarr2}) is reduced to that
of charged AdS black holes for $z=0$ . It is also notable that in the
extended phase space, the modified first law of thermodynamics is completely
in agreement with the modified Smarr relation and can be written as%
\begin{equation}
dM=TdS+VdP+\mathcal{C}dC_{+}.  \label{GenFirstLaw}
\end{equation}%
Moreover, we can calculate the modified volume in the Lifshitz like
spacetime as%
\begin{equation}
V=\left( \frac{\partial M}{\partial P}\right) _{S,C_{+}}=\frac{4\pi }{3}%
r_{+}^{2+b_{-}}.  \label{Volume}
\end{equation}%
where we see that the Lifshitz parameter modify the thermodynamic volume and
for $z=0$ one can recover the usual volume $V=\frac{4\pi r_{+}^{3}}{3}$.

Hereafter, since we desire to investigate the possible van der Waals phase
transition, we regard negative cosmological constant (positive $P$) and
spherical horizon ($S^{2}$) with volume $\omega _{1}=4\pi $. Regarding the
relation for the temperature (\ref{Temp2}) with the mentioned equation of
pressure (\ref{P}), we can obtain the equation of state $P=P(T,r_{+})$, as%
\begin{equation}
P=\frac{12\pi r_{+}\left( \frac{r_{+}}{r_{0}}\right) ^{-\frac{z}{2}%
}T-3C_{+}\left( b_{+}-b_{-}\right) r_{+}^{-b_{+}}-3Kb_{+}}{8\pi
r_{+}^{2}\left( 2+b_{+}\right) },  \label{pressure}
\end{equation}%
where $r_{+}$ is a function of the thermodynamic volume as indicated in Eq. (%
\ref{Volume}). Following Ref. \cite{Dolan}, one can find a relation between
the horizon radius and specific volume. In the geometric units, we can
obtain
\begin{equation}
r_{+}=\frac{v}{2}.  \label{vol}
\end{equation}

As a result, we can work with specific volume, or directly, the radius of
event horizon as volume representative.

Supposing the existence of critical behavior of the system in the $P-V$
isotherm diagram, one finds that the inflection point of $P-V$ diagram can
be interpreted as the critical point on the critical isotherm with the
following properties
\begin{equation}
\left( \frac{\partial P}{\partial r_{+}}\right) _{T}=\left( \frac{\partial
^{2}P}{\partial r_{+}^{2}}\right) _{T}=0.  \label{infel}
\end{equation}

Using Eq. (\ref{infel}) and the equation of state (\ref{pressure}), we can
calculate the critical parameters as
\begin{eqnarray}
T_{c} &=&\frac{2Kb_{+}b_{-}\xi ^{\frac{2-z}{2b_{+}}}}{\pi
(z+2)(2b_{+}+2-z)r_{0}^{\frac{z}{2}}},  \notag \\
r_{c} &=&\frac{v_{c}}{2}=\xi ^{\frac{-1}{b_{+}}},  \notag \\
P_{c} &=&-\frac{3K(z-2)b_{-}b_{+}}{8\pi (2+z)(2+b_{-})(2+b_{+})}\xi ^{\frac{2%
}{b_{+}}},  \label{Crit}
\end{eqnarray}

These relations lead us to obtain the following ratio
\begin{equation*}
\rho _{c}=\frac{P_{c}v_{c}}{T_{c}}=\frac{-3(z-2)(2b_{+}+2-z)r_{0}^{\frac{z}{2%
}}}{8(2+b_{-})(2+b_{+})}\xi ^{\frac{z}{2b_{+}}},
\end{equation*}%
where the constant $\xi $ is defined as%
\begin{equation*}
\xi =\frac{2K(z-2)b_{-}}{(2b_{+}+2-z)(2+b_{+})(b_{+}-b_{-})C_{+}}.
\end{equation*}%
It is easy to check that for $z=0$, the mentioned ratio reduces to the van
der Waals fluid, $\frac{3}{8}$.


\begin{figure}[tbp]
$%
\begin{array}{ccc}
\epsfxsize=5.3cm \epsffile{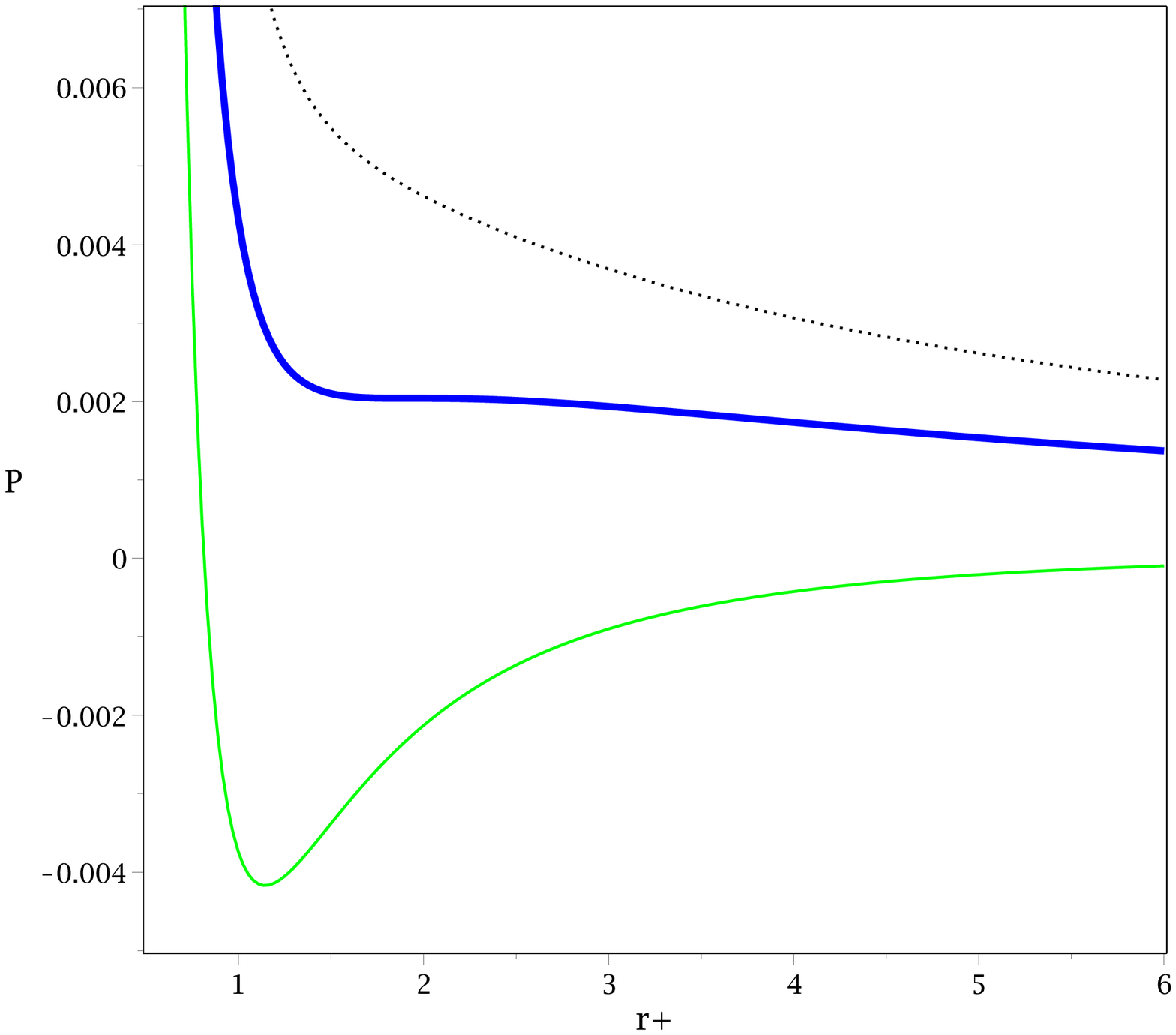} & \epsfxsize=5.9cm %
\epsffile{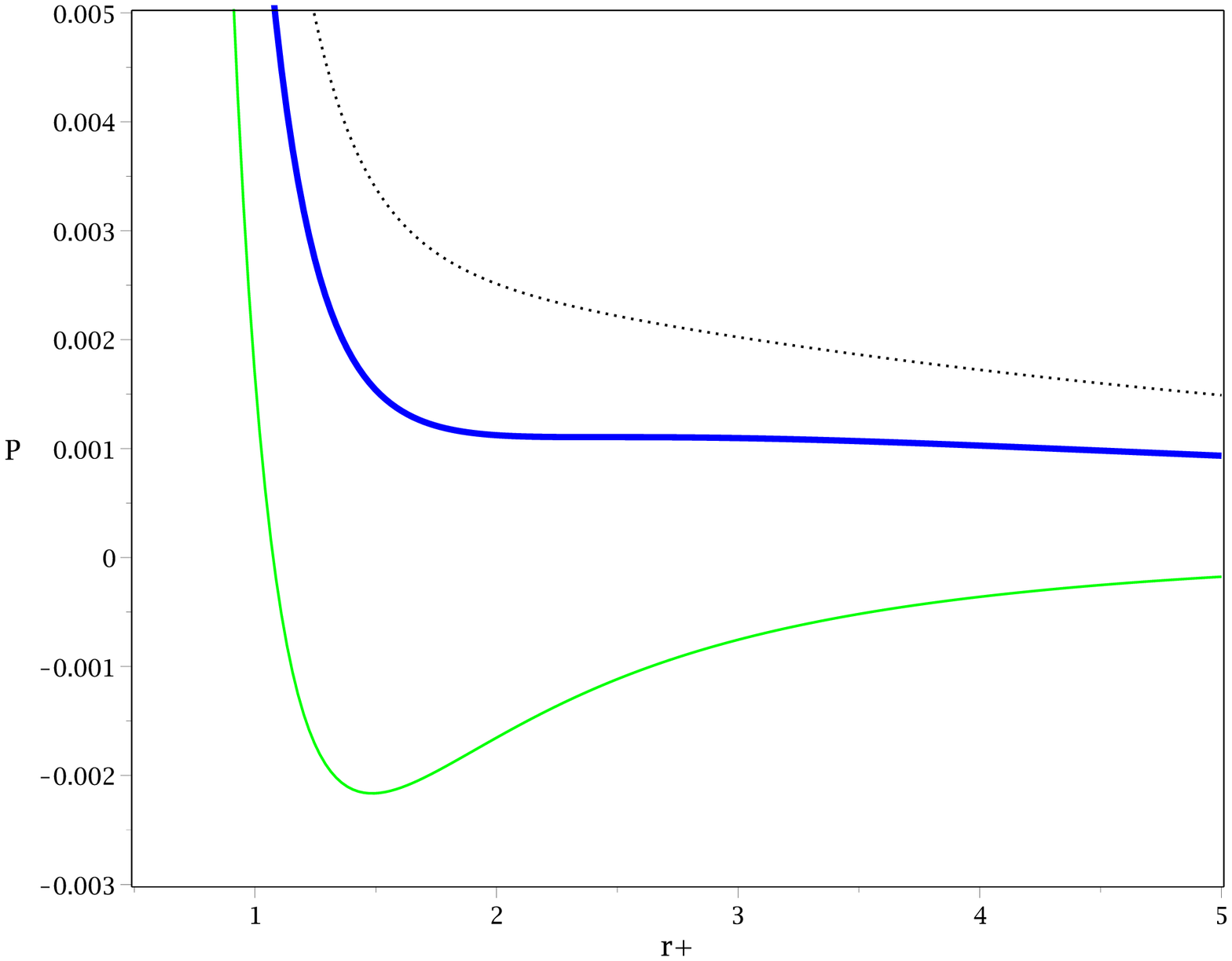} & \epsfxsize=5.5cm \epsffile{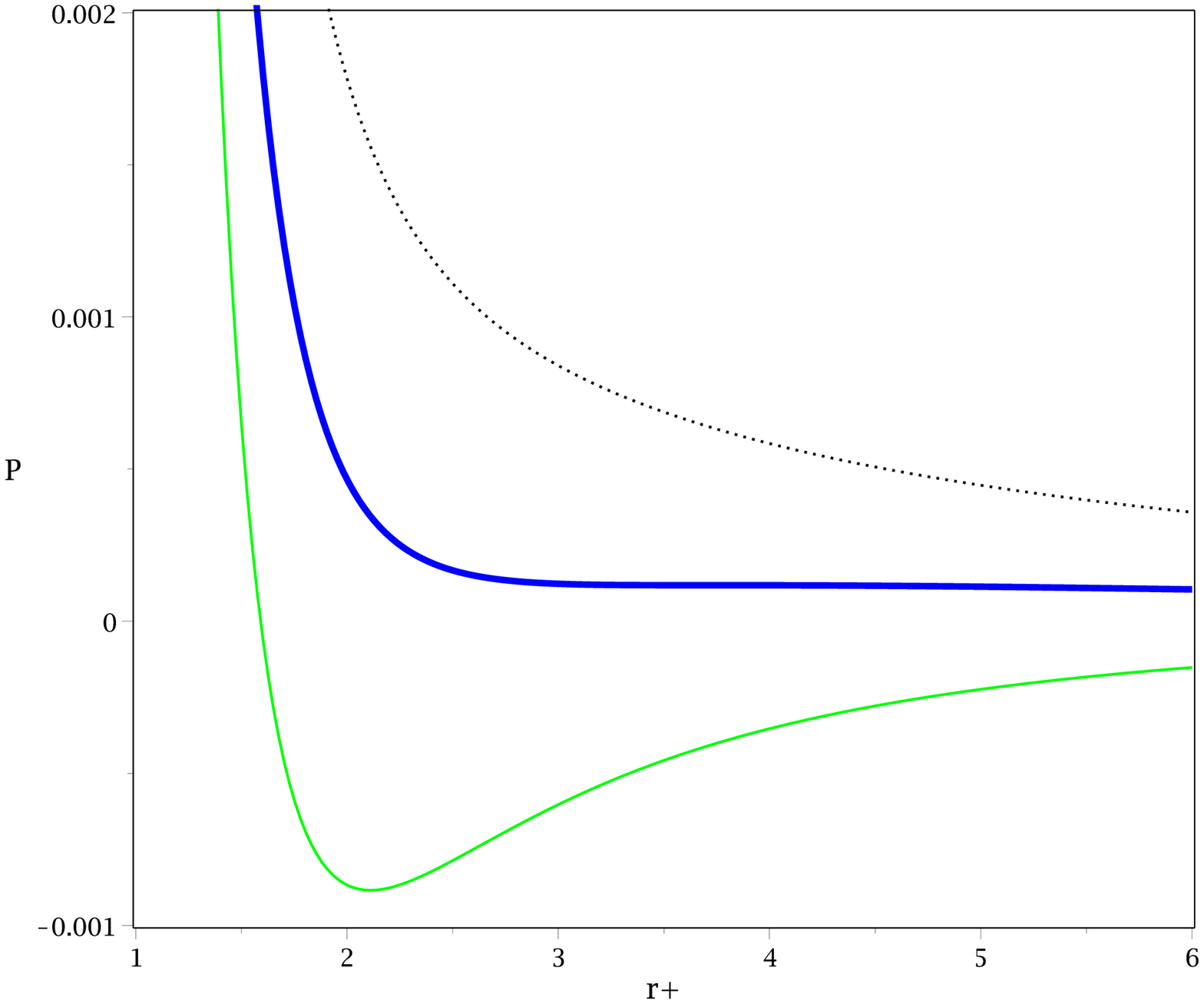}%
\end{array}
$%
\caption{ $P-r_{+}$ diagrams for $C_{+}=-1$, $k=1$, $r_{0}=1$, $T<T_{c}$
(green continuous line), $T=T_{c}$ (blue bold line) and $T>T_{c}$ (black
dashed line). "Note: $z=-0.1$ (left panel), $z=0$ (middle panel) and, $z=1$
(right panel)"}
\label{Fig-PV}
\end{figure}

On the other hand, thermodynamic behavior of a system can be investigated by
some thermodynamic potentials such as the free energy. It is known that the
enthalpy of black holes in the extended phase space is the total mass. The
reason is due to the fact that the cosmological constant is no longer a
fixed parameter but a thermodynamical one. Due to modification in
interpretation of the total mass of the black holes in extended phase space,
the Gibbs free energy is given by%
\begin{equation}
G=M-TS,  \label{Gibbs}
\end{equation}%
or equevalently its value per unit volume $\omega _{k}$ is
\begin{eqnarray}
G &=&\frac{K}{2}r_{+}^{b_{-}}\left( 1-\frac{b_{+}}{1+b_{-}-\frac{z}{2}}%
\right) +\frac{4}{3}\pi Pr_{+}^{2+b_{-}}\left( 1-\frac{2+b_{+}}{1+b_{-}-%
\frac{z}{2}}\right)  \notag \\
&&\left. -\frac{C_{+}}{2r_{+}^{b_{+}-b_{-}}}\left( 1+\frac{b_{+}-b_{-}}{%
1+b_{-}-\frac{z}{2}}\right) \right] .  \label{Gibbs2}
\end{eqnarray}

In order to investigate thermal behavior of the obtained black holes, we
have plotted $P-r_{+}$ isotherms and $G-T$ diagrams. The presence of both
swallow-tail characteristic in $G-T$ diagrams and the inflection point in $%
P-r_{+}$ diagrams indicate that our system undergoes a first order phase
transition. Generally speaking, the van der Waals like phase transition
between two different phases is characterized by a swallow-tail shape in $%
G-T $ diagrams and the inflection point in $P-r_{+}$ plots. In our case,
such a behavior represents a phase transition between small and large black
holes.

\begin{figure}[tbp]
$%
\begin{array}{ccc}
\epsfxsize=5.4cm \epsffile{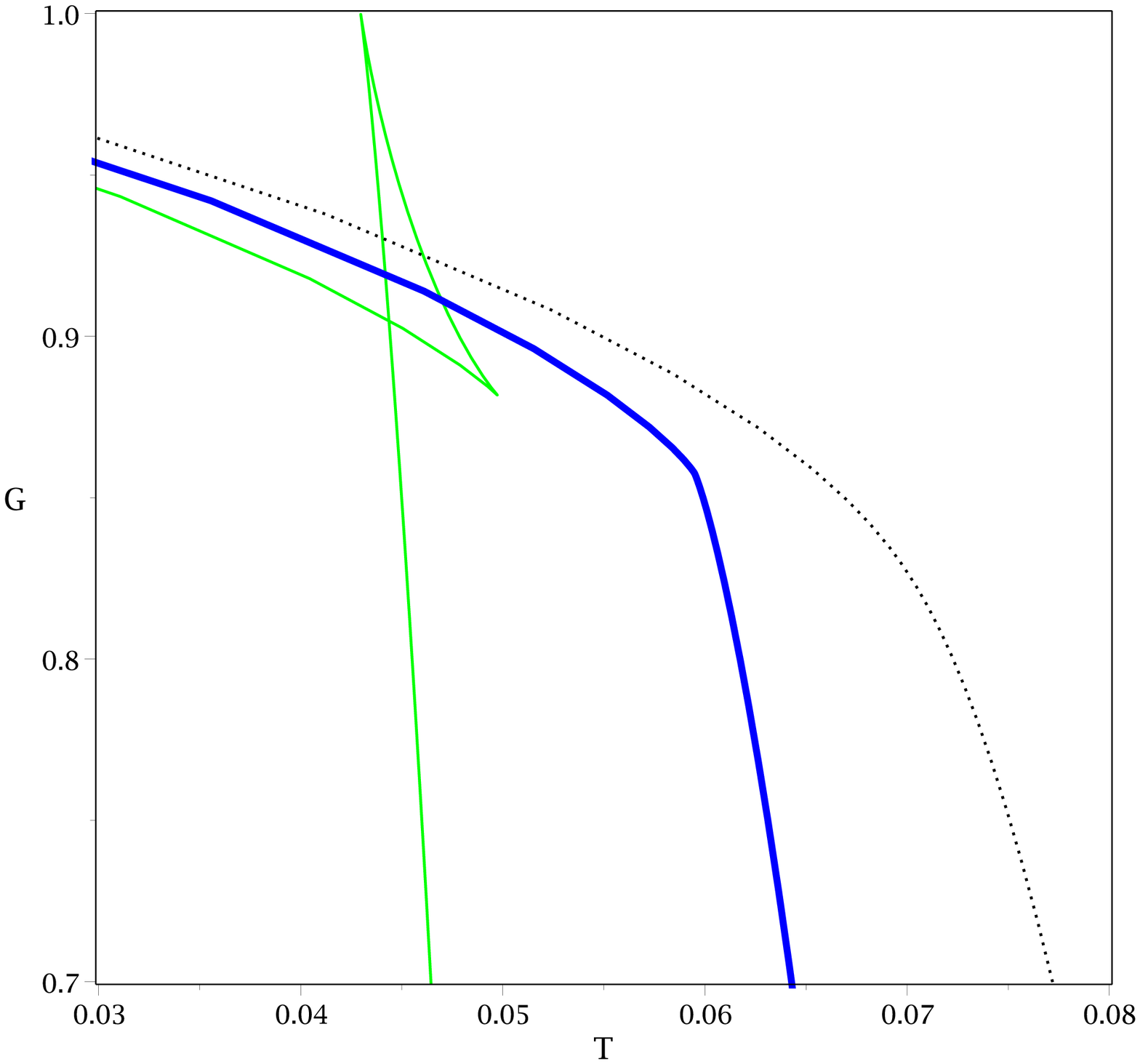} & \epsfxsize=5.7cm %
\epsffile{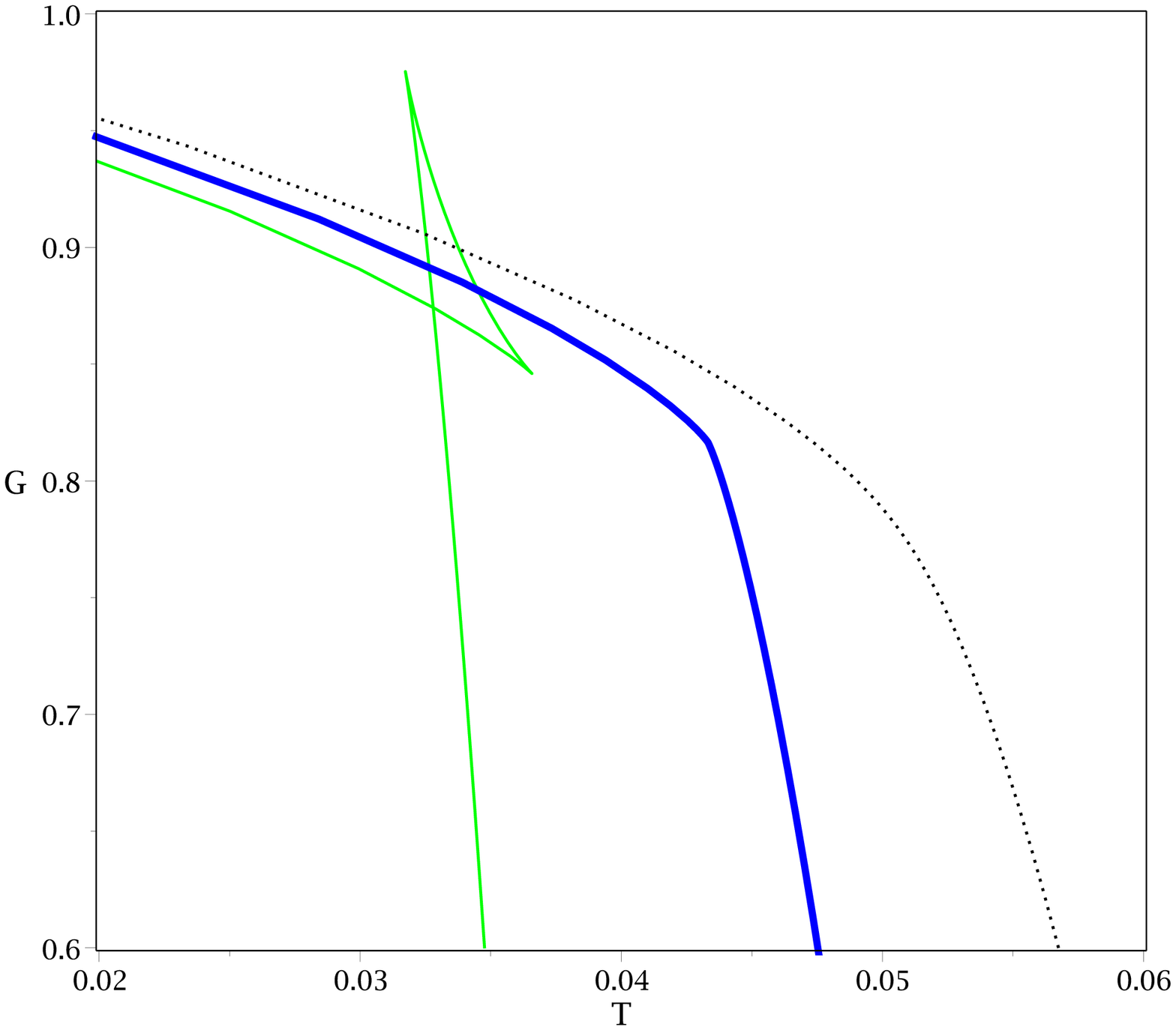} & \epsfxsize=6cm \epsffile{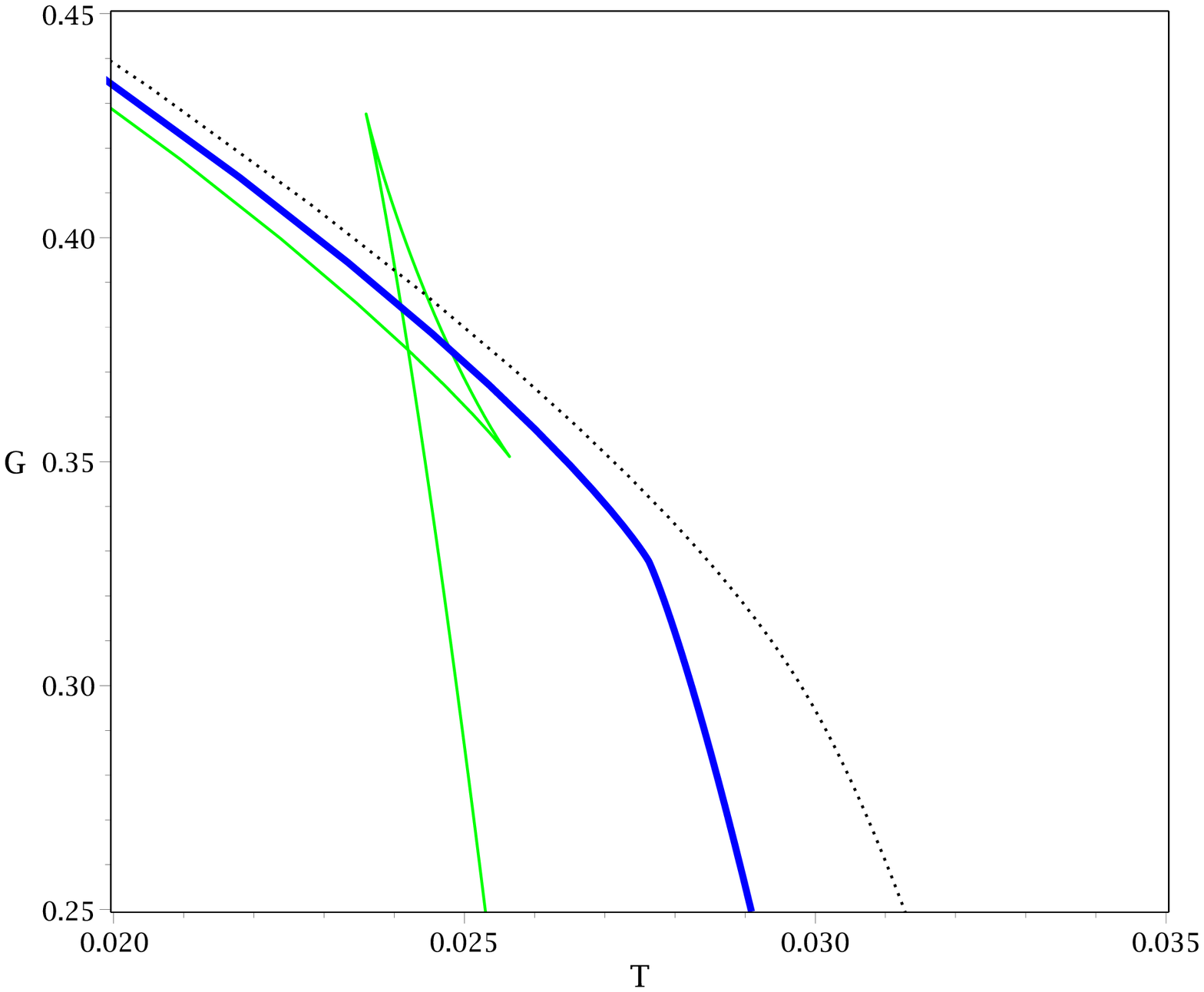}%
\end{array}
$%
\caption{ Gibbs free energy versus temperature for $C_{+}=-1$, $k=1$, $%
r_{0}=1$, $P<P_{c}$ (green continuous line), $P=P_{c}$ (blue bold line) and $%
P>P_{c}$ (black dashed line). "Note: $z=-0.1$ (left panel), $z=0$ (middle
panel) and $z=1$ (right panel)"}
\label{Fig-GT}
\end{figure}

\begin{table}[]
\caption{critical quantities for $k=1$, $r_{0}=1$ and $C_{+}=-1$}
\label{Tabel 1}\centering
\begin{tabular}{|c|c|c|c|c|}
\hline\hline
z & $T_c$ & $v_c=2r_{c}$ & $P_c$ & $\rho_{c}=\frac{P_c v_c}{T_c}$ \\
\hline\hline
-0.1 & 0.0595 & 3.85 & 0.0061 & 0.397 \\ \hline
0.0 & 0.043 & 4.90 & 0.0033 & 0.375 \\ \hline
0.1 & 0.036 & 5.53 & 0.0023 & 0.343 \\ \hline
0.5 & 0.028 & 6.72 & 0.0009 & 0.210 \\ \hline
1.0 & 0.027 & 7.37 & 0.0003 & 0.095 \\ \hline
1.5 & 0.029 & 8.17 & 0.0001 & 0.030 \\ \hline\hline
\end{tabular}%
\end{table}

\begin{figure}[tbp]
$%
\begin{array}{cc}
\epsfxsize=7cm \epsffile{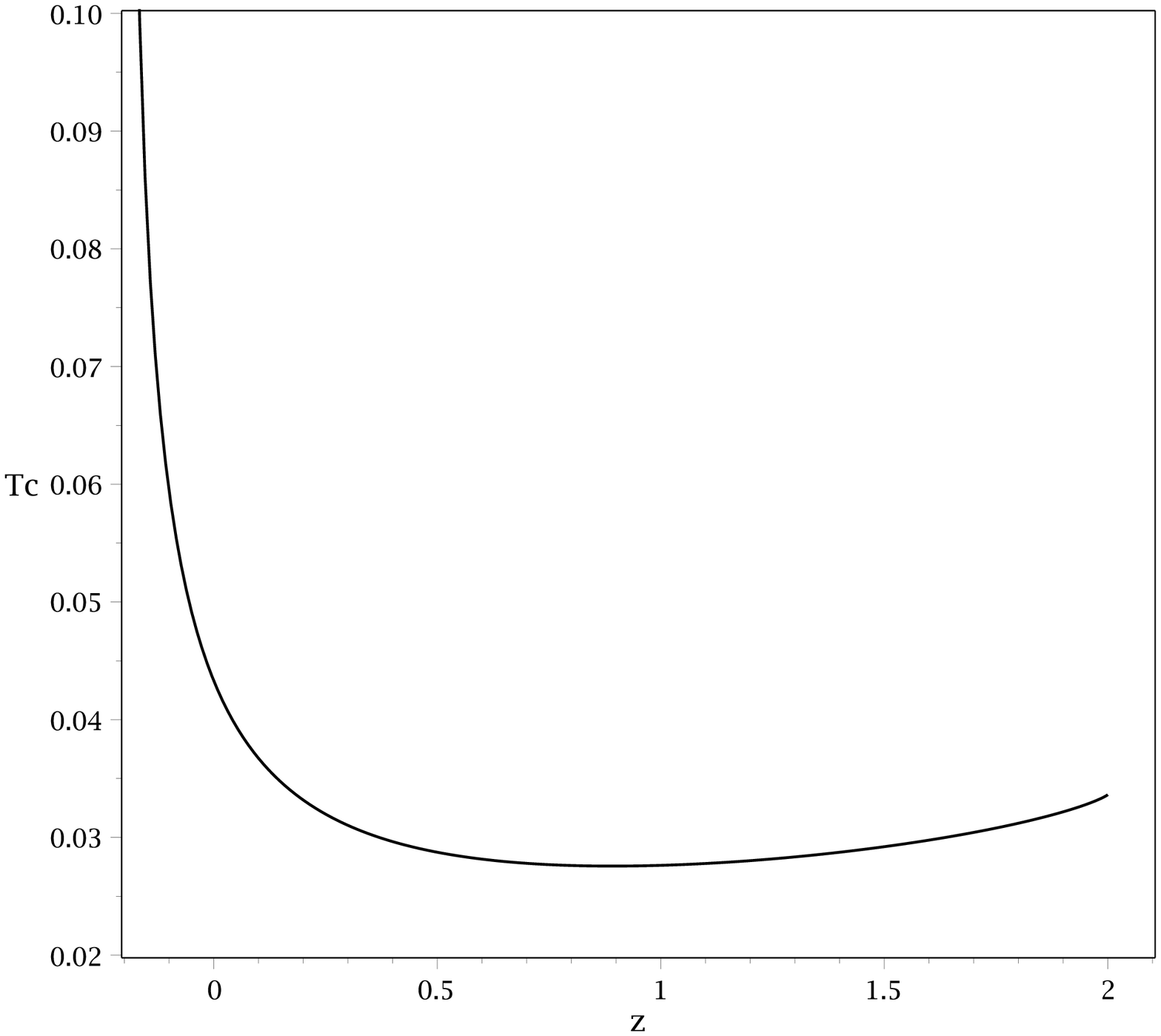} & \epsfxsize=7cm \epsffile{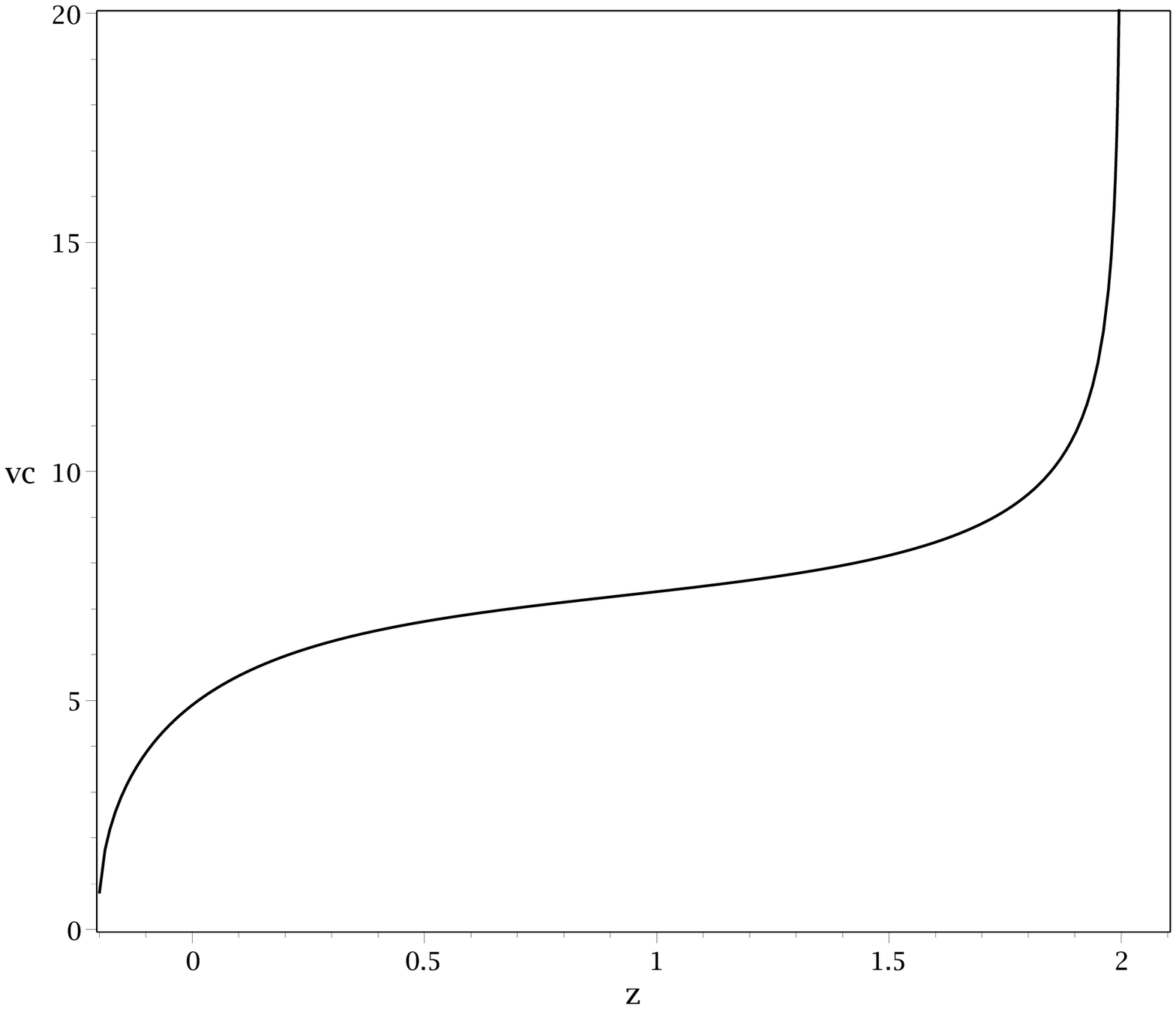} \\
\epsfxsize=7cm \epsffile{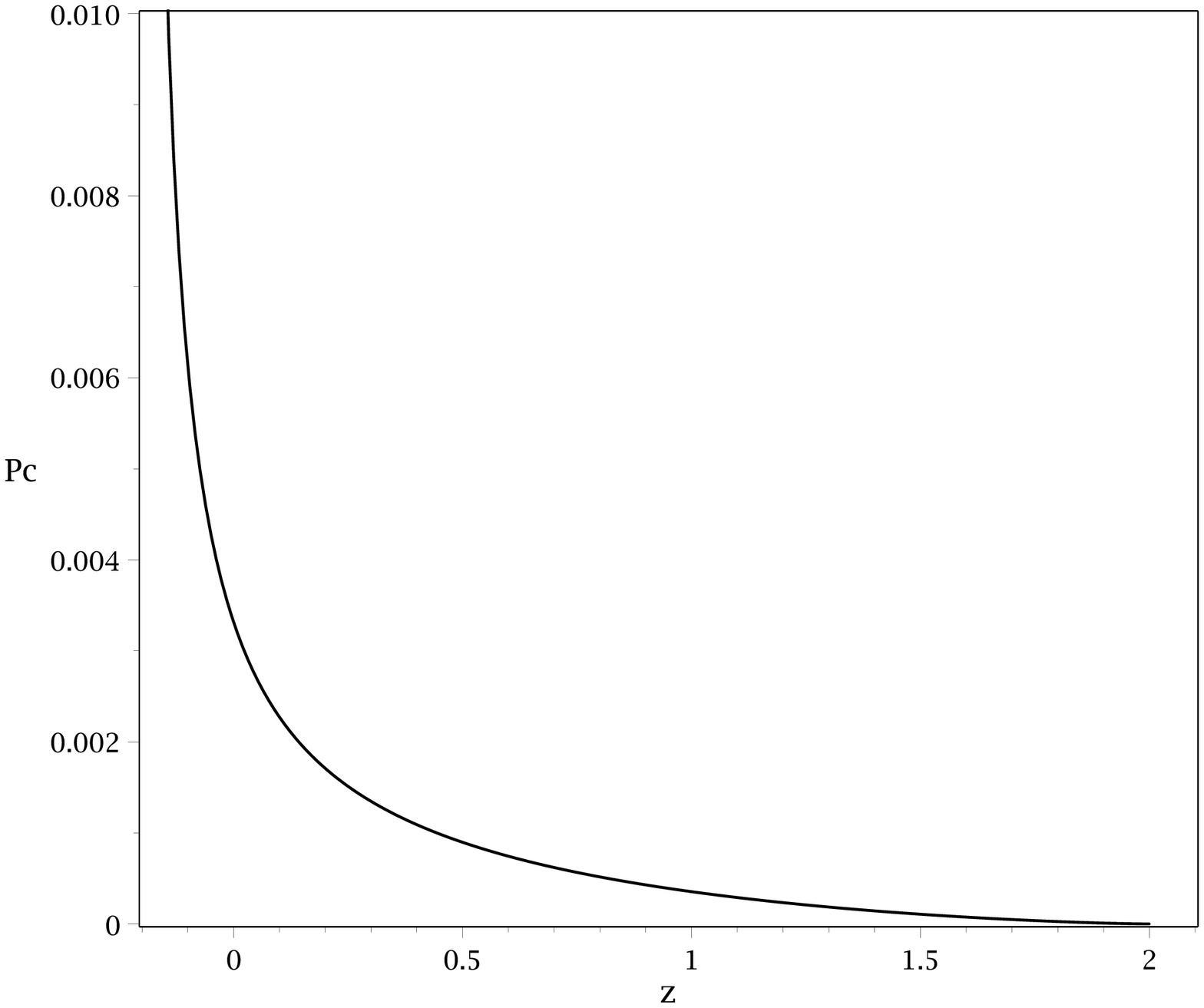} & \epsfxsize=7cm \epsffile{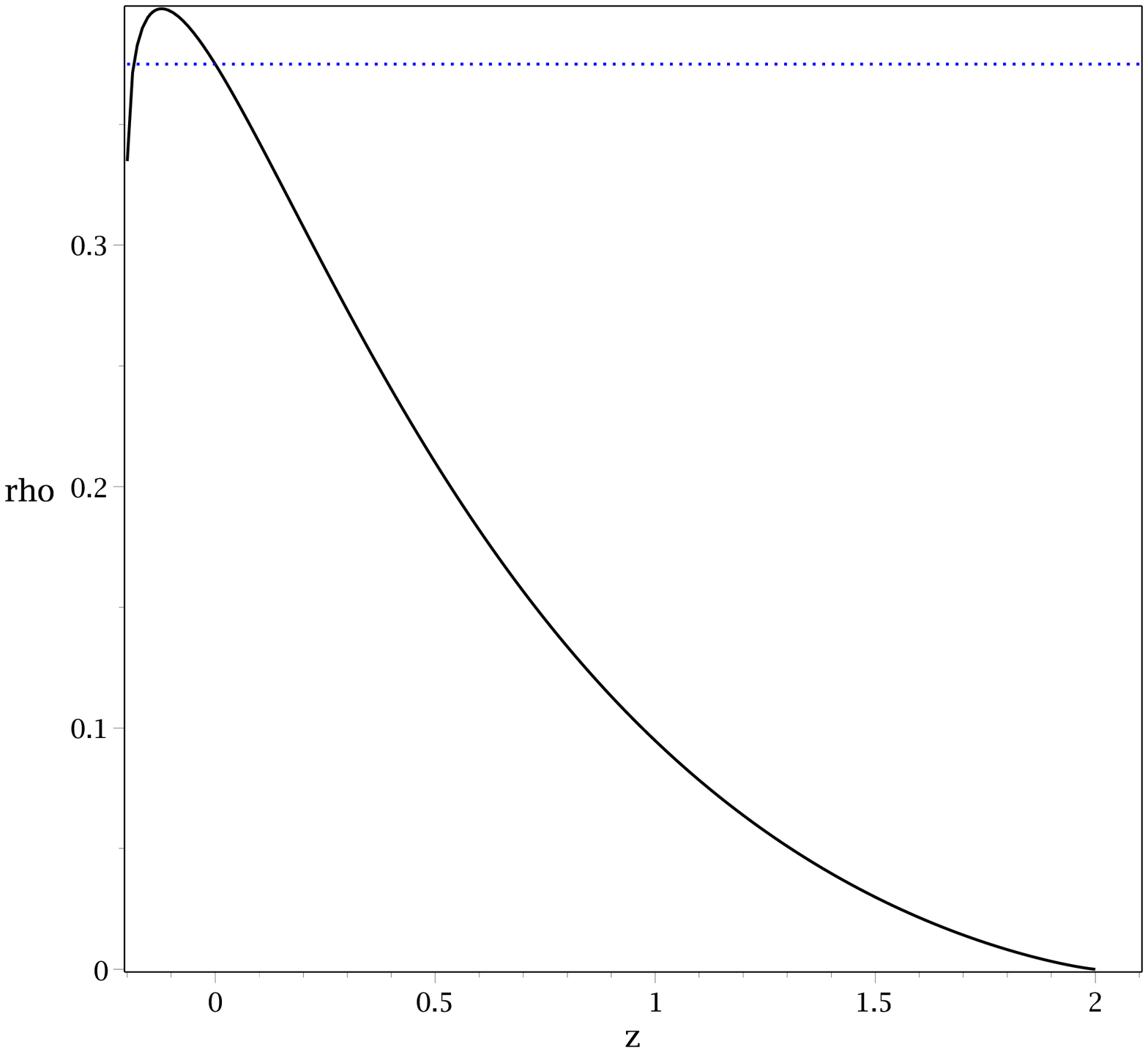}%
\end{array}
$%
\caption{ Critical quantities versus $z$ for $C_{+}=-1$, $k=1$ and $r_{0}=1$%
. \textbf{Left-up panel:} $T_c$ versus $z$, \textbf{right-up panel:} $v_c$
versus $z$, \textbf{Left-down panel:} $P_c$ versus $z$, \textbf{right-down
panel:} black continuous line $\protect\rho_c$ versus $z$ and blue dotted
line is $\protect\rho_c=\frac{3}{8}$ }
\label{Critical}
\end{figure}

Now, we desire to investigate thermal stability of the obtained black hole
solutions. In order to discuss thermal stability of a black hole, one can
calculate the heat capacity and discuss its sign (positivity/negativity)
through its roots and divergencies. Basically, the conditions regarding
thermal stability of black holes could be attained by studying the sign of
heat capacity. Regardless of the values of parameters in the theory,
positivity of the heat capacity ensures thermal stability of the solutions,
whereas its negativity is considered to be an unstable state. Another
advantage of investigating the heat capacity is the relation of its
divergencies with phase-transition interpretation. Since we are working in
the extended phase space, the heat capacity is given by
\begin{equation}
C_{Q,P}=T{\left( \frac{\partial S}{\partial T}\right) _{Q,P}}=T\frac{\left(
\frac{\partial S}{\partial r_{+}}\right) _{Q,P}}{\left( \frac{\partial T}{%
\partial r_{+}}\right) _{Q,P}}.  \label{CQ}
\end{equation}

By using the equations of temperature (\ref{Temp2}) and entropy (\ref{TU}),
we can obtain the heat capacity of black holes as%
\begin{equation}
C_{Q,P}=\frac{4\pi r_{+}^{1+b_{-}}\chi }{\left( \frac{r_{+}}{r_{0}}\right) ^{%
\frac{z}{2}}},  \label{CQ1}
\end{equation}%
where%
\begin{equation*}
\chi =\frac{8\pi
P(2+b_{-})r_{+}^{2+b_{+}}+3Kb_{-}r_{+}^{b_{+}}-3C_{+}(b_{-}-b_{+})}{8\pi
P(2+b_{-})(z+2)r_{+}^{2+b_{+}}+3Kb_{-}(z-2)r_{+}^{b_{+}}-3C_{+}(b_{-}-b_{+})(z-2b_{+}-2)%
}
\end{equation*}

\begin{figure}[tbp]
$%
\begin{array}{ccc}
\epsfxsize=5.5cm \epsffile{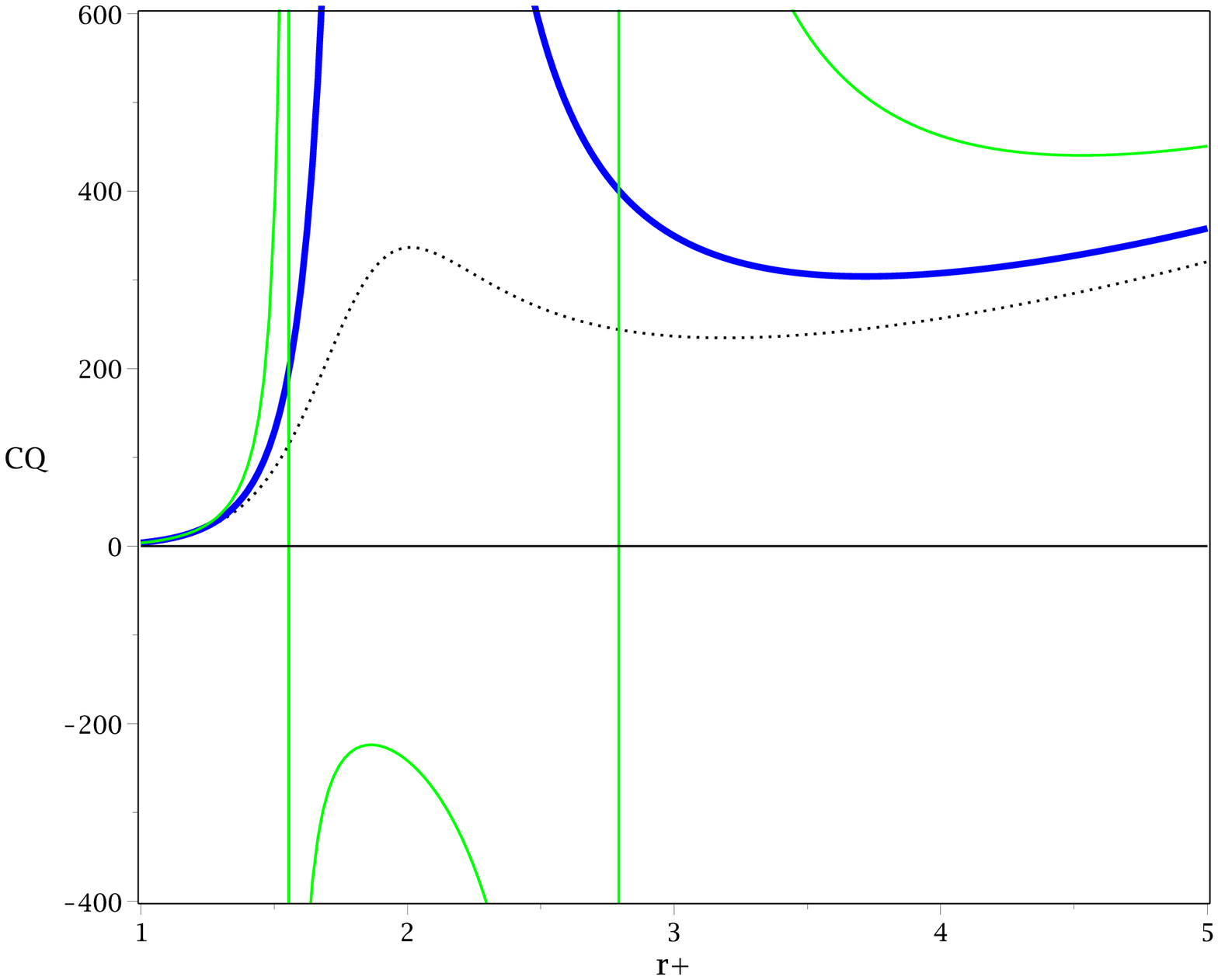} & \epsfxsize=5.5cm %
\epsffile{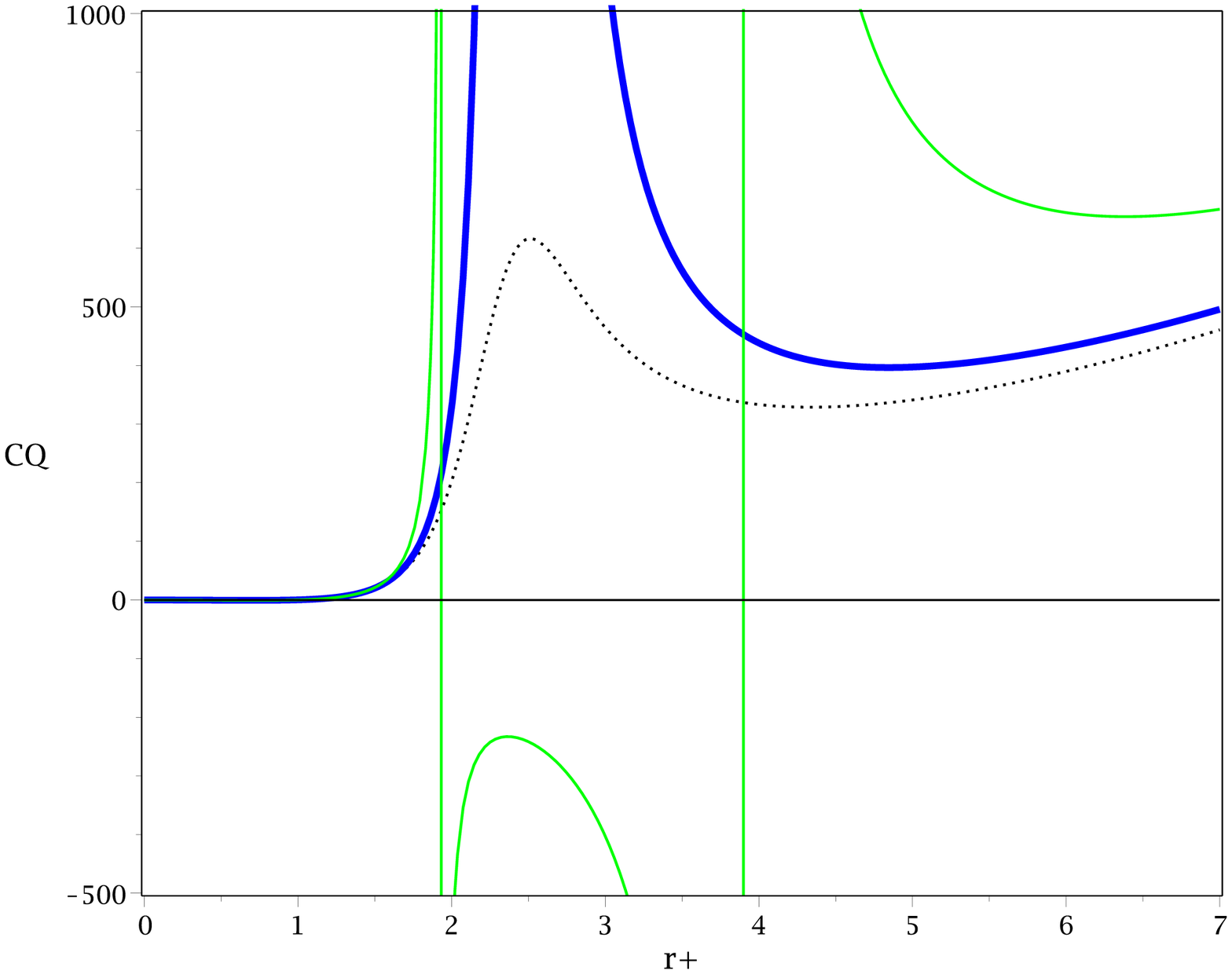} & \epsfxsize=5.5cm \epsffile{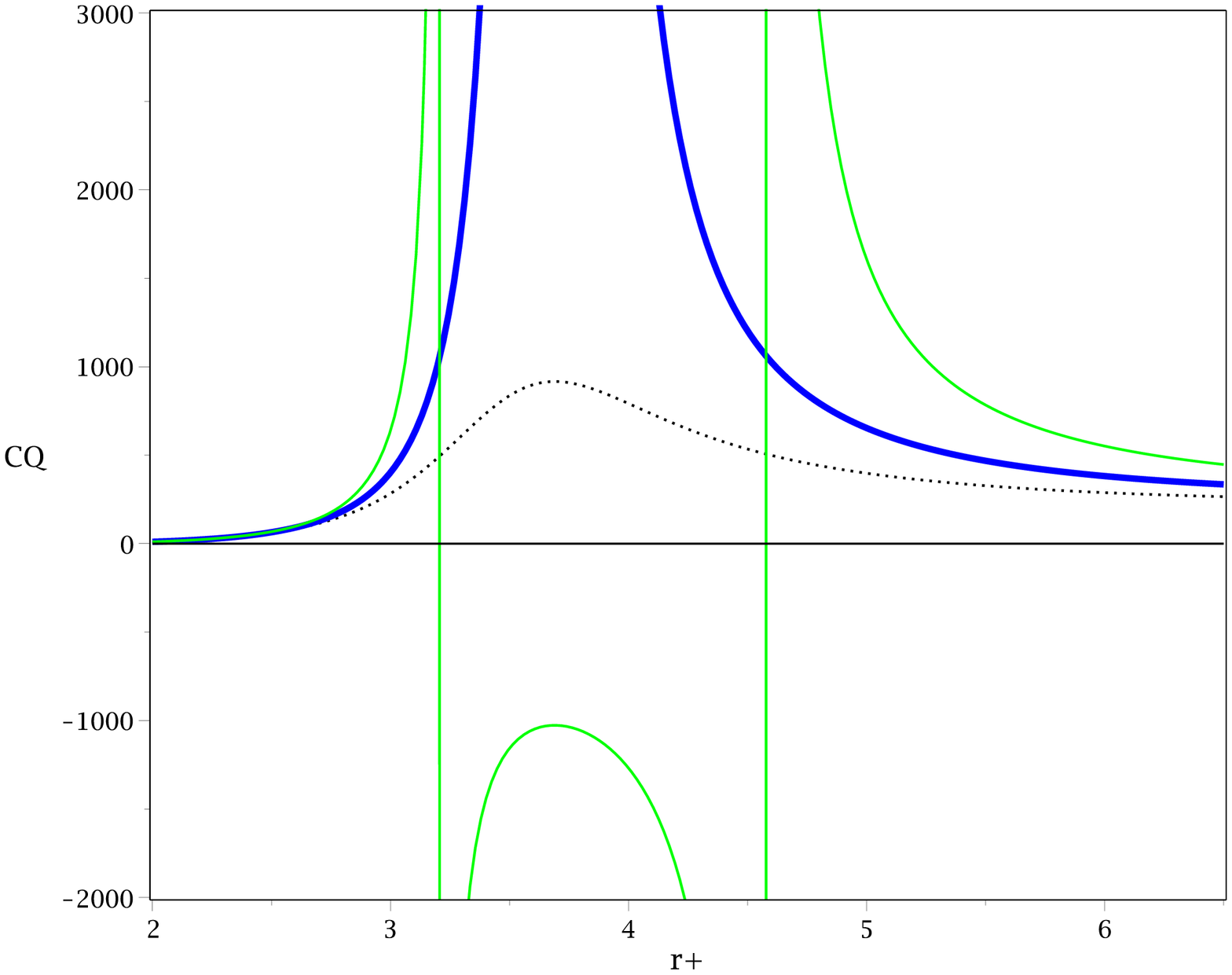}%
\end{array}
$%
\caption{ Heat Capacity versus $r_{+}$ for $C_{+}=-1$, $k=1$, $r_{0}=1$, $%
P<P_{c}$ (green continuous line), $P=P_{c}$ (blue bold line) and $P>P_{c}$
(black dashed line). "Note: $z=-0.1$ (left panel), $z=0$ (middle panel) and $%
z=1$ (right panel)"}
\label{Fig-CG}
\end{figure}

\subsection{Results and discussion}

\label{Discussion}

Regarding Figs. \ref{Fig-PV}, \ref{Fig-GT} and \ref{Fig-CG}, one can find
that for $T<T_c$ (in $P-r_{+}$ diagrams) or $P<P_{c}$ (in $G-T$ and $%
C_{Q,P}-r_{+}$ diagrams) there is a van der Waals like phase transition for
the obtained black hole solutions. The critical behavior is indicated as
blue-bold line in the mentioned diagrams. As we expected the results of all
diagrams are consistent.

Strictly speaking, regarding isotherm $P-r_{+}$ diagrams, we find that for $%
T>T_{c}$ an ideal gas behavior with no phase transition is observed. The
critical isotherm is plotted for $T=T_{c}$ with an inflection point at $%
P=P_{c}, r_{+}=r_{c}$. For $T<T_{c}$ a van der Waals like shape is appeared.

Taking into account $G-T$ diagrams, one can find a smooth curve for $P>P_c$,
a continuously curve which is not differentiable at a point with $P=P_c$ and
$T=T_c$, and a swallow-tail shape for $P<P_c$, which is the characteristic
of a phase transition in $G-T$ diagrams.

In addition, having a look at the heat capacity diagrams, we find that there
is thermally stable black holes with positive definite heat capacity for $%
P>P_c$, while for the critical case, $P=P_c$, there is only one divergence
point in the heat capacity diagrams, in which the positive sign of $C_{Q,P}$
does not change around the singularity. Finally for $P<P_c$, we observe two
divergence points, in which the heat capacity is negative between them, and
therefore, there is a phase transition between the mentioned two divergence
points.

Having a glance at the critical quantities, one can find different
behaviors. According to Fig. \ref{Critical}, we find that $P_c$ ($v_{c}$) is
a decreasing (an increasing) function of $z$, while there is a minimum
(maximum) for $T_{c}$ ($\rho_{c}$). Therefore, in order to have a universal
ratio such as the van der Waals fluid, one can obtain $z \simeq -0.1853$, in
addition to $z=0$. It is also worth mentioning that for $z \longrightarrow
2^{-}$, both $P_c$ and $\rho_{c}$ vanish and criticality is disappeared. For
$z>2$, all critical values are complex and one cannot obtain real valued
critical quantities. All the mentioned calculations are done for the
spherical horizon, $k=1$. Since some of the critical quantities depend on
the topological factor ($k$), linearly, one cannot find a nonzero critical
quantity for black holes with flat horizon. But for hyperbolic horizon, one
may look for the possible van der Waals like behavior, the same happens in
massive gravity \cite{PRDrapid}. We postpone this interesting subject to
appendix.

\section{Conclusion}

In this paper, we have studied thermodynamic behavior of topological black
hole solutions with Lifshitz-like spacetime. We have worked in a special
class of $F(R)$ gravity models with constant Ricci scalar in which satisfies
two simultaneous conditions, $F(R_{0})=0$ and $F_{R}=0$. We have shown that
although these solutions are asymptotically AdS with an effective
cosmological constant for nonzero Ricci scalar, for vanishing Ricci scalar
obtained solutions are not asymptotically flat unless for vanishing Lifshitz
parameter.

We also calculated thermodynamic quantities and found that the solutions of
this special class of modified gravity do not undergo the usual entropy and
mass that reported in $F(R)$ gravity black holes, since $F_{R}=0$. We have
calculated the entropy by using the first law of thermodynamics and found
that it reduces to area law for vanishing Lifshitz parameter, $z=0$.

In addition, we investigate the phase transition in the extended phase space
thermodynamics by considering the cosmological constant (which is
proportional to the constant Ricci scalar) as a thermodynamical pressure for
the spherical horizon black holes. We have studied the van der Waals like
behavior by investigating three diagrams: isothermal pressure-volume,
isobaric Gibbs free energy-temperature, and isobaric (isocharge) heat
capacity-horizon radius. We have found that all the mentioned diagrams have
consistent results to show three cases: completely stable state without any
phase transition, critical behavior in the critical diagrams, and existence
of a van der Waals like phase transition. We have shown that the Lifshitz
parameter has an important role for the values of critical quantities.

We also extended our calculations to the case of hyperbolic horizon black
holes and showed that a van der Waals like behavior can be observed only for
negative temperature. In other words, for $k=-1$, although all physical
quantities are positive, temperature is negative which may explain based on
a quantum behavior of the black holes. Since it is a new result in the
context of $F(R)$ gravity, it will be interesting to work on its nature by
calculation of temperature based on statistical mechanics and simulate it
with a quantum system with negative temperature.

\section{Appendix A: van der Waals like behavior for $k=-1$}

\begin{figure}[h]
$%
\begin{array}{ccc}
\epsfxsize=5.4cm \epsffile{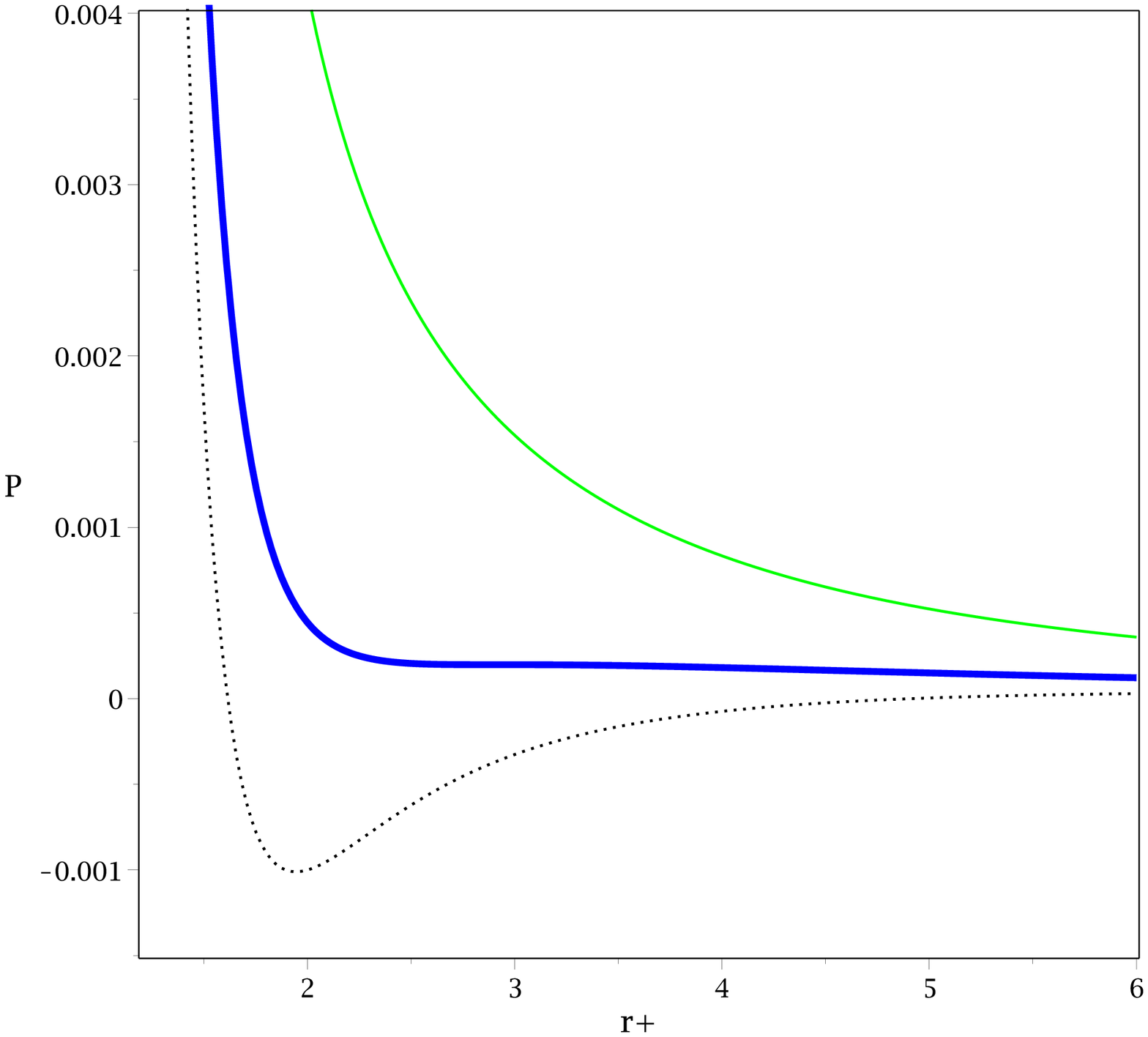} & \epsfxsize=5.7cm %
\epsffile{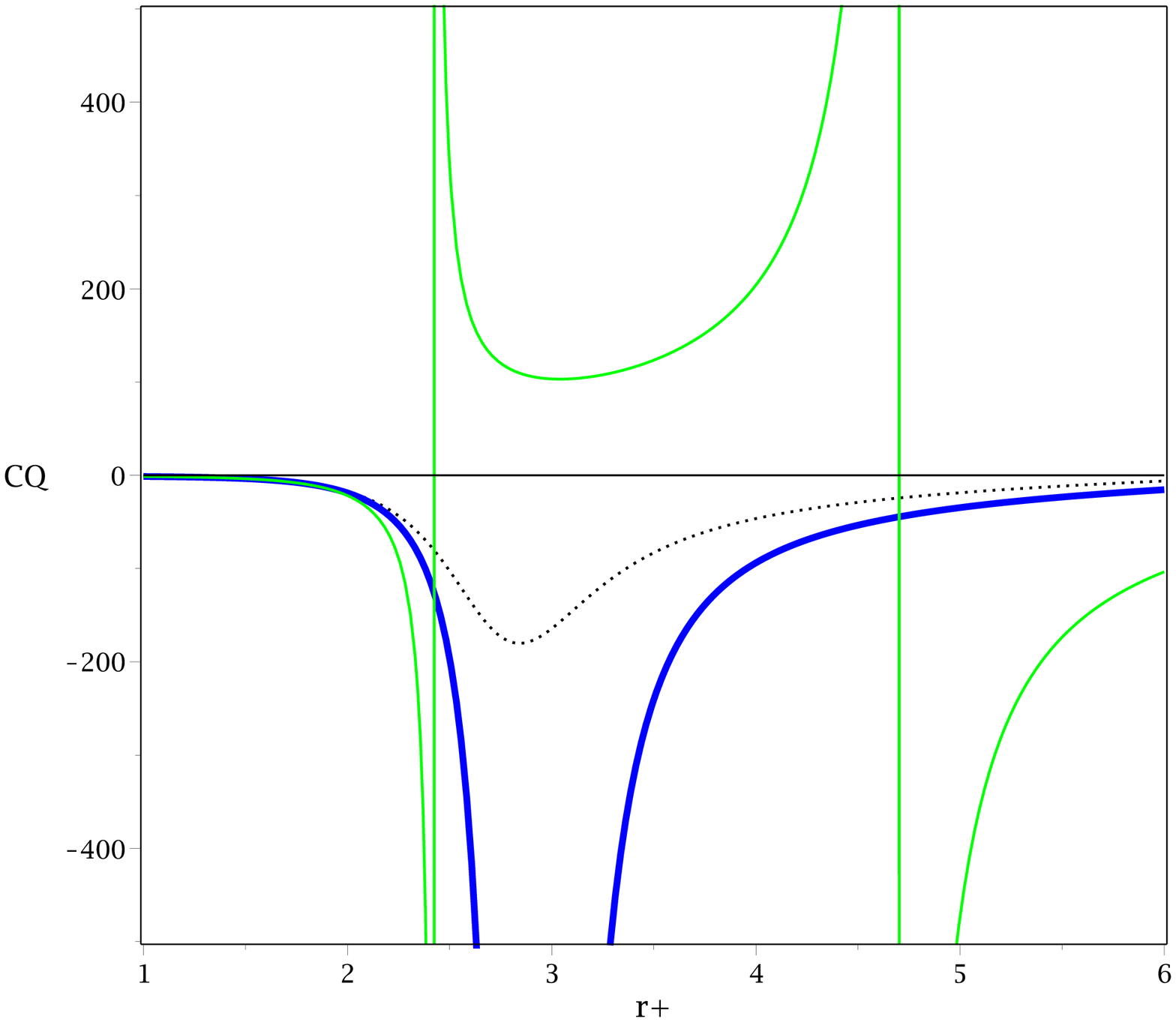} & \epsfxsize=6cm \epsffile{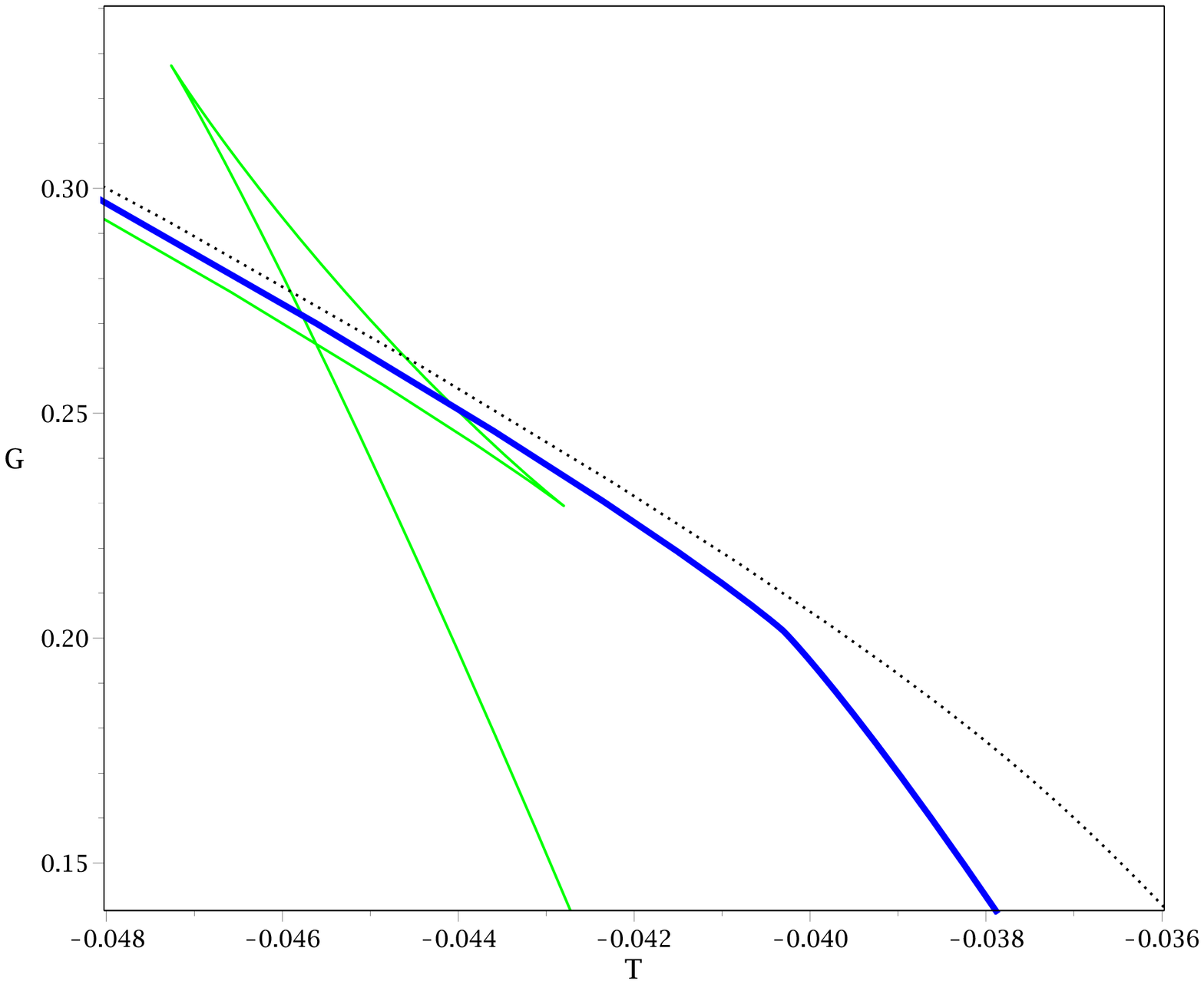}%
\end{array}
$%
\caption{ Diagrams of $P-r_{+}$ (left), $C_{Q,P}-r_{+}$ (middle) and $G-T$
(right) for $C_{+}=-1$, $k=-1$, $r_{0}=1$, $z=3$.}
\label{Fig-negative-k}
\end{figure}

Here, we are going to examine the possible van der Waals like behavior for
the black hole solutions with hyperbolic horizon ($k=-1$) . According to Eq.
(\ref{Crit}), Fig. \ref{Fig-negative-k} and table II, we find that a van der
Waals behavior for hyperbolic horizon is observed. It is notable that
although the critical volume and pressure are positive, the critical
temperature is negative. In other words, all physical quantities such as
Gibbs free energy, volume and pressure are positive in this case, but the
same as those of nuclear spins system in an external magnetic field and
population inversion in laser, temperature is negative. In this regard,
although the swallow-tail shape in $G-T$ diagram and van der Waals behavior
in $P-r_{+}$ figure make sense, the interpretation of phase transition
(between two stable states) based on the (positivity of) heat capacity
should change. Since negative temperature of physical system is a quantum
mechanical behavior (it is not observed in classical systems), it will be
interesting to investigate the thermodynamic behavior of the mentioned black
hole solutions with the statistical mechanical approach.

\begin{table}[h]
\caption{critical quantities for $k=-1$, $r_{0}=1$ and $C_{+}=-1$}
\label{Tabel 2}\centering
\begin{tabular}{|c|c|c|c|c|}
\hline\hline
z & $T_c$ & $v_c=2r_{c}$ & $P_c$ & $\rho_{c}=\frac{P_c v_c}{T_c}$ \\
\hline\hline
2.5 & -0.0382 & 3.48 & 0.00009 & -0.008 \\ \hline
3.0 & -0.0403 & 2.91 & 0.00020 & -0.014 \\ \hline
3.5 & -0.0417 & 2.61 & 0.00030 & -0.019 \\ \hline
4.0 & -0.0426 & 2.41 & 0.00038 & -0.022 \\ \hline
4.5 & -0.0433 & 2.27 & 0.00044 & -0.023 \\ \hline
5.0 & -0.0438 & 2.16 & 0.00049 & -0.024 \\ \hline\hline
\end{tabular}%
\end{table}

\section{Appendix B: dynamical stability}

Here, we consider a massless scalar perturbation in the background of the
black hole spacetime and obtain the quasi-normal frequencies (QNFs) by using
the third order WKB approximation. The WKB approximation was first applied
to the problem of scattering around black holes \cite{Schutz}, and then
extended to the third order \cite{IyerWill}. This method can be used for an
effective potential that forms a barrier potential and takes constant values
at the event horizon and cosmological horizon. In addition, we concentrate
our attention to the fixed values of free parameters of metric function
throughout the text.

The equation of motion for a massless scalar field is given by
\begin{equation}
\square \Phi =0.  \label{W}
\end{equation}

If we consider modes as
\begin{equation}
\Phi \left( t,r,\theta ,\varphi \right) =\frac{1}{r}\Psi \left( r\right)
Y_{l,m}\left( \theta ,\varphi \right) e^{-i\omega t},
\end{equation}%
where $Y_{l,m}\left( \theta ,\varphi \right) $ is the spherical harmonics,
the equation of motion (\ref{W})\ reduces to the following wave equation
\begin{equation}
\left[ \partial _{x}^{2}+\omega ^{2}-V_{l}\left( x\right) \right] \Psi
_{l}\left( x\right) =0,  \label{Weq}
\end{equation}%
where $x$ is the tortoise coordinate
\begin{equation}
dx=\frac{dr}{e^{\alpha \left( r\right) }B(r)},  \label{tortoise}
\end{equation}%
and the effective potential $V_{l}\left( x\right) $ is given by%
\begin{equation}
V_{l}\left( x\right) =e^{2\alpha \left( r\right) }B(r)\left[ \frac{l\left(
l+1\right) }{r^{2}}+\frac{B(r)\alpha ^{\prime }\left( r\right) }{r}+\frac{%
B^{\prime }\left( r\right) }{r}\right] ,  \label{EP}
\end{equation}%
where $l$ is the angular quantum number. One can obtain the QNMs of this
perturbation by considering proper boundary conditions as follows
\begin{equation}
\begin{array}{c}
\Psi _{l}\left( r\right) \sim e^{-i\omega x}\ \ \ \ \ \ as\ \ \ \ \ \
x\rightarrow -\infty \ (r\rightarrow r_{e}), \\
\Psi _{l}\left( r\right) \sim e^{i\omega x}\ \ \ \ \ \ \ as\ \ \ \ \ \ \ \ \
x\rightarrow \infty \ (r\rightarrow r_{c}),%
\end{array}
\label{bc}
\end{equation}%
which means that no wave comes from the event horizon and cosmological
horizon ($r_{e}$ is the event horizon and $r_{c}$ is the cosmological
horizon). We should consider theses boundary conditions to obtain the QNFs.

The third order WKB formula is given by
\begin{equation}
\frac{i\left( \omega ^{2}-V_{0}\right) }{\sqrt{-2V_{0}^{\prime \prime }}}%
-\Lambda _{2}-\Lambda _{3}=n+\frac{1}{2};\ \ \ \ \ \ n=0,1,2,...,
\end{equation}%
where $V_{0}$ is the value of effective potential at its local maximum, $n$
is the overtone number, and the correction terms $\Lambda _{2}$ and $\Lambda
_{3}$ are given in \cite{IyerWill}. The results are given in the tables $III$
and $IV$. The real (imaginary) part of the frequencies decreases (increases)
as the overtone number increases, but the angular quantum number has
opposite behavior. On the other hand, increasing in $z$ leads to decreasing
both the real and imaginary parts of the frequencies.

\begin{table}[h]
\caption{QNMs ($\protect\omega _{r}-i\protect\omega _{i}$) for $C_{+}=0.25$,
$C_{-}=0.25$, $k=1$, $z=0$, $R_{0}=1$,\ and $r_{0}=1$.}
\label{Tabel 3}\centering
\begin{tabular}{|c|c|c|c|c|c|}
\hline\hline
$n$ & $l=1$ & $l=2$ & $l=3$ &  &  \\ \hline\hline
$0$ & $0.9514-0.4385i$ & $1.622-0.4218i$ & $2.281-0.4173i$ &  &  \\ \hline
$1$ & $-$ & $1.510-1.294i$ & $2.197-1.267i$ &  &  \\ \hline
$2$ & $-$ & $-$ & $2.051-2.144i$ &  &  \\ \hline\hline
\end{tabular}%
\vspace{0.2cm}
\end{table}

\begin{table}[h]
\caption{The fundamental QNMs ($\protect\omega _{r}-i\protect\omega _{i}$)
for $C_{+}=0.25 $, $C_{-}=0.25$, $k=1$, $R_{0}=1$,\ and $r_{0}=1$.}
\label{Tabel 4}\centering
\begin{tabular}{|c|c|c|c|c|c|}
\hline\hline
$z$ & $l=1$ & $l=2$ & $l=3$ &  &  \\ \hline\hline
$-0.1$ & $1.035-0.4775i$ & $1.758-0.4598i$ & $2.471-0.4553i $ &  &  \\ \hline
$0.5$ & $0.6881-0.2859i$ & $1.180-0.2733i$ & $1.661-0.2696i $ &  &  \\ \hline
$1.0$ & $0.5409-0.1941i$ & $0.9247-0.1827i$ & $1.304-0.1797i$ &  &  \\ \hline
$1.5$ & $0.4295-0.1361i$ & $0.7396-0.1285i$ & $1.045-0.1268i$ &  &  \\ \hline
$2.0$ & $0.3247-0.09380i$ & $0.5670-0.09095i$ & $0.8038-0.09034i$ &  &  \\
\hline
$2.5$ & $0.1978-0.05376i$ & $0.3490-0.05331i$ & $0.4959-0.05322i$ &  &  \\
\hline\hline
\end{tabular}%
\vspace{0.2cm}
\end{table}

\begin{figure}[tbp]
$%
\begin{array}{cccc}
\epsfxsize=5cm \epsffile{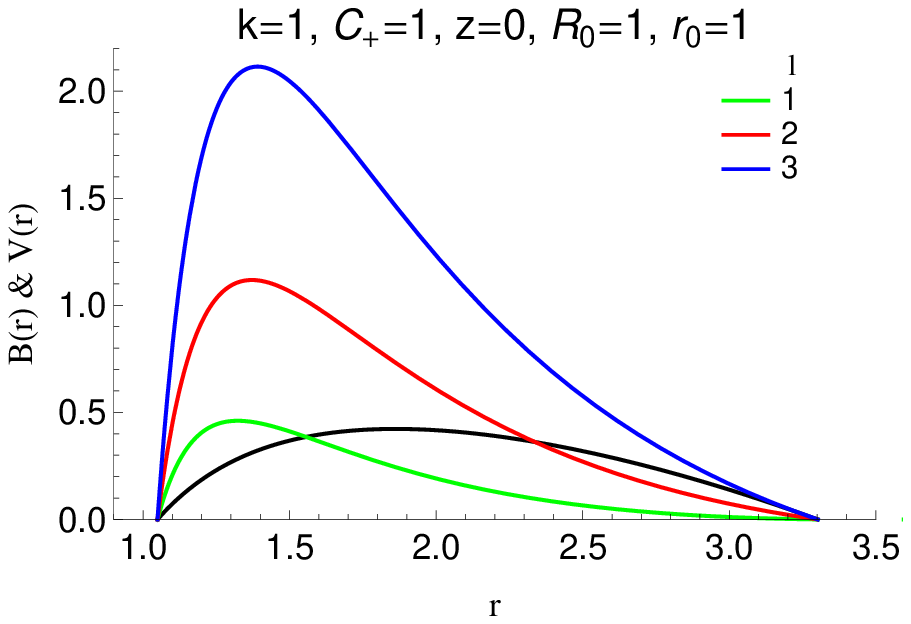} & \epsfxsize=5cm \epsffile{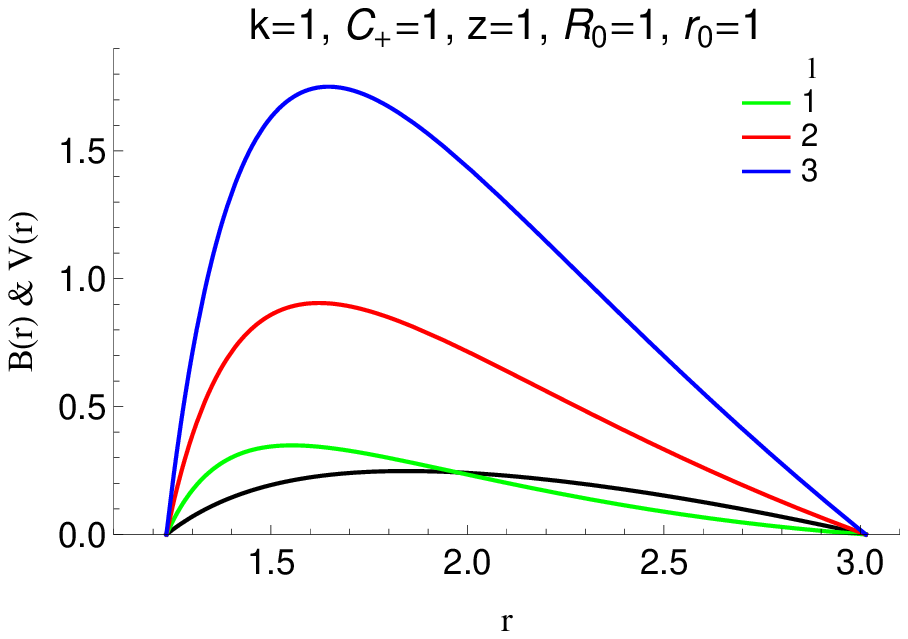} & %
\epsfxsize=5cm \epsffile{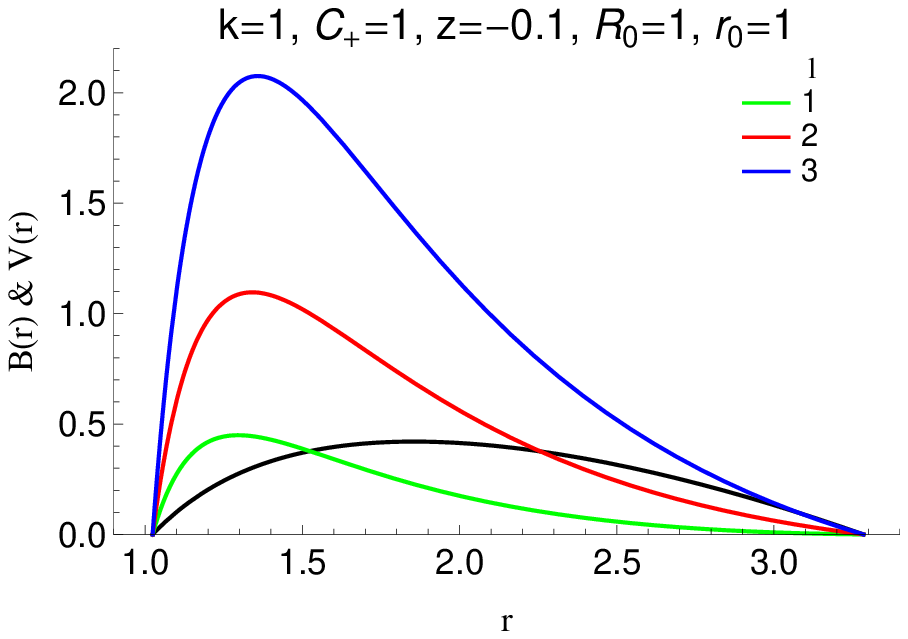} &
\end{array}
$%
\caption{The effective potential (colored lines) and metric function (black
line) versus the radial coordinate $r$.}
\label{Pot}
\end{figure}

In the case of dynamical stability, the figure \ref{Pot} shows the behavior
of the effective potential in which this potential forms a barrier
potential. Since the effective potentials are positive everywhere, then $%
\int_{-\infty }^{+\infty }V_{l}\left( x\right) dx>0$ and the obtained black
hole solutions are dynamically stable under massless scalar perturbations
\cite{Simon}. Therefore, it is possible to find the dynamically stable black
holes by using the obtained black hole solutions.

\section{Appendix C: non-constant Ricci scalar}

Now, we give a discussion regarding an arbitrary function of Ricci scalar ($%
R=R(r)$). Considering the metric ansatz (\ref{Metric2}) with an arbitrary
Ricci scalar, one finds the following exact solutions
\begin{equation}
B(r)=K-\frac{C_{\pm }}{r^{b_{\pm }}}\pm \frac{(2+b_{-})(2+b_{+})}{%
(b_{+}-b_{-})r^{b_{\pm }}}\int r^{1+b_{\pm }}\lambda (r)dr,  \label{B2}
\end{equation}%
where $K$ and $b_{\pm }$ are introduced before, and following Eq. (\ref%
{lambda}), we define $\lambda (r)=\frac{2R(r)}{z^{2}+8z+24}$. It is
straightforward to show that Eq. (\ref{B2}) reduces to (\ref{B(r)}) for
constant Ricci scalar, $\lambda (r)=\lambda $.

In order to obtain the temperature, we follow Eq. (\ref{Temp}). After some
manipulations, we obtain
\begin{equation}
T=\left\{ \frac{Kb_{-}r_{+}^{b_{+}}-C_{+}\left( b_{+}-b_{-}\right)
-(2+b_{+})(2+b_{-})\Theta _{+}}{4\pi r_{+}^{1+b_{+}}}\right\} \left( \frac{%
r_{+}}{r_{0}}\right) ^{\frac{z}{2}},
\end{equation}%
in which $\Theta _{+}=\int^{r=r_{+}}r^{1+b_{+}}\lambda (r)dr$.

As we mentioned before, we have to calculate the entropy via the first law
of thermodynamics, (\ref{first}). As one expects, the new relation of
entropy is the same as that of reported in Eq. (\ref{TU}).

Now, one may looking for possible phase transition. To do so, we have to
specify the functional form of the Ricci scalar. Specifying the Ricci
scalar, one can define a suitable dynamical pressure and investigate
critical behavior of the solutions. Since it is straightforward, we abandon
the presentation of a specific example, for the sake of brevity.

\begin{acknowledgements}
We are grateful to the anonymous referees for the insightful
comments and suggestions, which have allowed us to improve this
paper significantly. SHH wishes to thank Shiraz University
Research Council. This work has been supported financially by the
Research Institute for Astronomy and Astrophysics of Maragha,
Iran.
\end{acknowledgements}

\end{document}